\documentclass[%
reprint,
superscriptaddress,
nobibnotes,
amsmath,
amssymb,
aps,
prl,
]{revtex4-2}

\usepackage{graphicx}
\usepackage{dcolumn}
\usepackage{bm}
\usepackage[mathlines]{lineno}

\begin{document}

\title{Pseudospin Transverse Localization of Light in an Optical Disordered Spin-Glass Phase}

\author{Shani Izhak}
\altaffiliation[Corresponding author: ]{shaniizhak@mail.tau.ac.il}
\affiliation{
School of Electrical Engineering, Iby and Aladar Fleischman Faculty of Engineering, Tel Aviv University, Tel Aviv 69978, Israel}
\author{Aviv Karnieli}
\affiliation{
Raymond and Beverly Sackler School of Physics and Astronomy, Tel Aviv University, Tel Aviv 69978, Israel}
\affiliation{
Department of Applied Physics, Stanford University, Stanford, CA 94305, USA}
\author{Ofir Yesharim}
\affiliation{
School of Electrical Engineering, Iby and Aladar Fleischman Faculty of Engineering, Tel Aviv University, Tel Aviv 69978, Israel}
\author{Shai Tsesses}
\affiliation{
Andrew and Erna Viterbi Department of Electrical Engineering, Technion--Israel Institute of Technology, Haifa 3200003, Israel}
\affiliation{
Department of Physics, Center for Ultracold Atoms and Research Laboratory of Electronics, Massachusetts Institute of Technology, Cambridge, MA 02139, USA}
\author{Ady Arie}
\affiliation{
School of Electrical Engineering, Iby and Aladar Fleischman Faculty of Engineering, Tel Aviv University, Tel Aviv 69978, Israel}

\date{\today}

\begin{abstract}
Localization phenomena during transport are typically driven by disordered \textit{scalar} potentials. Here, we predict a universal pseudospin localization phenomenon induced by a disordered \textit{vectorial} potential and demonstrate it experimentally in an optical analogue of a classical disordered spin-glass magnetic phase. In our system, a transverse disorder in the second-order nonlinear coupling of a nonlinear photonic crystal causes the idler-signal light beam, representing the pseudospin current, to become localized in the transverse plane. This effect depends strongly on the nonlinear coupling strength, controlled by the optical pump power, revealing its inherently nonlinear and non-perturbative nature. Furthermore, this phenomenon is marked by decaying Rabi oscillations between the idler and signal fields, linked to the disorder properties, suggesting an accompanied longitudinal decoherence effect. Our findings offer deep insights into spin transport in disordered magnetic textures and open avenues for exploring complex magnetic phases and phase transitions using nonlinear optics.
\end{abstract}

\maketitle

Magnetic phases of matter and spin transport within them are fundamental topics with broad implications, particularly in spintronics \cite{doi:10.1126/science.1065389}. The spin-glass phase, a disordered system where spins are aligned randomly due to frustrated interactions, is especially intriguing \cite{S_F_Edwards_1975, PhysRevLett.52.1156, mezard1987spin, RevModPhys.65.829, doi:10.1073/pnas.0601120103}, attracting interest across diverse scientific fields \cite{S_F_Edwards_1975, PhysRevLett.50.1946, PhysRevLett.52.1156, mezard1987spin, RevModPhys.65.829, doi:10.1073/pnas.0601120103, PhysRevX.5.031040, Fan2023}, and earning Giorgio Parisi the 2021 Nobel Prize in Physics \cite{PhysRevLett.50.1946, mezard1987spin, doi:10.1073/pnas.0601120103, NobelPrize2021}. However, despite extensive theoretical work, experimentally probing spin currents in disordered magnetization textures like in spin-glass phases, remains challenging \cite{RevModPhys.58.801, GOODENOUGH2002297, PhysRevLett.99.086401, Appelbaum2007, PhysRevLett.106.247203, 10.1063/5.0046083, Han2020, Chen2022}. Precise control over spin currents and the study of phase transition dynamics in such complex systems, are still limited, underscoring the need for systems that offer greater control over spin transport in disordered magnetic phases.

In disordered systems, phenomena like weak localization \cite{PhysRevLett.55.2692, PhysRevLett.55.2696}, and Anderson localization \cite{PhysRev.109.1492, 10.1063/1.3206091} are predicted to occur. In Anderson localization, a disordered potential halts the diffusion of an electron wavefunction, leading to spatial localization with exponentially decaying tails. Lightwaves can also exhibit localization \cite{Segev2013}. This has been observed in both completely random media \cite{AGRANOVICH1988378, PhysRevLett.55.2692, PhysRevLett.55.2696, PhysRevLett.71.3947, Wiersma1997, M_V_Berry_1997, Chabanov2000, Faez:09, PhysRevLett.96.063904, PhysRevA.108.053501, Yamilov2023}, and disordered periodic media in the original sense of the Anderson model \cite{Schwartz2007, PhysRevLett.100.013906, Karbasi:12, Boguslawski:13, PhysRevLett.112.193902, PhysRevB.100.140202, 10.1063/1.5142161, doi:10.1126/sciadv.abn7769}, where localization arises from disordered \textit{scalar} potentials. However, the study of localization in disordered \textit{vectorial} potentials, such as those found in spin-glass magnetization textures—where disorder can be adjusted in both magnitude and orientation—has remained elusive. Additionally, the interplay between disorder and nonlinearity, although explored in optics, continues to pose open questions \cite{PhysRevLett.69.1807, Schwartz2007, PhysRevLett.100.013906, Fishman_2012, Segev2013, PhysRevLett.112.193902, Mafi:15, PhysRevLett.121.233901}.

Recently, spin transport in magnetic materials has been emulated using a sum-frequency generation (SFG) process within nonlinear photonic crystals (NLPCs) \cite{https://doi.org/10.1002/lpor.201300107, PhysRevA.94.023805, PhysRevLett.120.053901, Karnieli:18, Karnieli2021, Karnieli2021_FoP, Yesharim2022, Izhak:24}. In this approach, the amplitudes of the idler and signal fields define a pseudospin, whereas the nonlinear coupling—dependent on the pump field and the crystal's second-order susceptibility—acts as a synthetic magnetization texture, $\textbf{M(r)}$. The ability to manipulate the synthetic magnetization texture, either by spatially shaping the pump field \cite{Karnieli2021, Karnieli2021_FoP, Yesharim2022}, or by spatially modulating the NLPC \cite{PhysRevLett.120.053901, Karnieli2021, Karnieli2021_FoP, Liu2023, Izhak:24}, creates a versatile platform for studying spin transport phenomena in various 2D magnetization textures using light.

\begin{figure*}[ht]
\includegraphics[width=\textwidth]{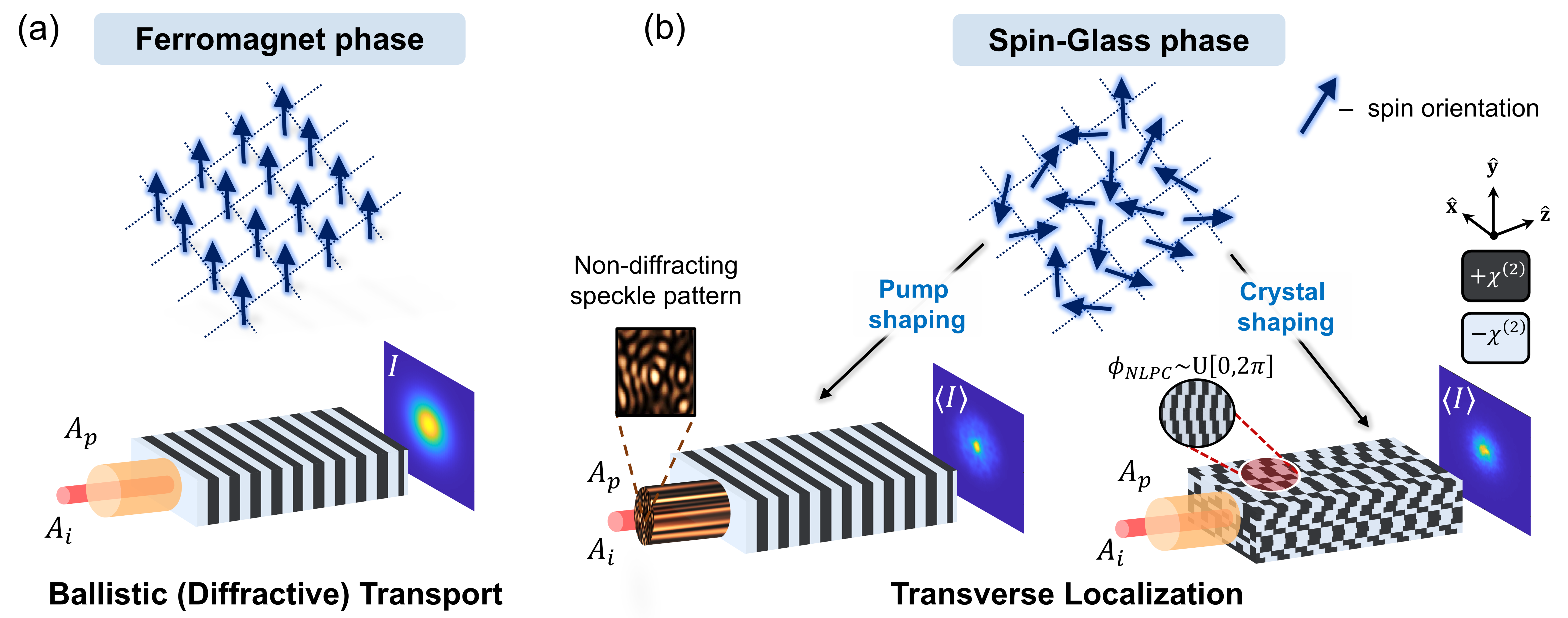}
\caption{\label{fig:figure1}\textbf{Pseudospin transverse localization of light.} (a) A ferromagnet phase (top) where the arrows represent the local magnetization, and its optical analogue scheme (bottom): Gaussian pump and idler beams impinge on a 1D-modulated NLPC, producing a diffracting Gaussian idler-signal beam. (b) A disordered spin-glass phase (top) and its optical analogue schemes (bottom): (left) a non-diffracting speckled pump and Gaussian idler beams impinge on a 1D-modulated NLPC, and (right) Gaussian pump and idler beams impinge on a 3D-modulated NLPC with its modulation phase is sampled from a uniform distribution within $\left[0, 2\pi\right]$. In both cases, the ensemble-averaged idler-signal output intensity (averaged over multiple disorder realizations) is transversely localized due to synthetic transverse disorder.}
\end{figure*}

In this letter, we leverage this analogy to study spin transport in a spin-glass magnetic phase using an all-optical platform. By manipulating the nonlinear interaction parameters, we introduce a disordered transverse synthetic magnetization pattern with randomly oriented pseudospins, emulating a 2D disordered spin-glass. Our numerical and experimental results show that, unlike in the optical ordered ferromagnet analogue, where the idler-signal pseudospin exhibits ballistic broadening, disorder in the synthetic magnetization leads to transverse localization of both idler and signal fields. Additionally, we show that the localization strength is highly dependent on the nonlinear coupling and reveal a temporal decoherence effect manifesting as decaying Rabi oscillations. This novel pseudospin transverse localization phenomenon, driven by vectorial disorder, opens avenues for studying complex magnetic phases and their dynamical transitions using light.

We consider an SFG process with pump, idler, and signal fields at frequencies $\omega_{p}, \omega_{i}$, and $\omega_{s}$, where $\omega_{s}=\omega_{p} + \omega_{i}$ due to energy conservation \cite{10.5555/1817101}. Assuming a quasi-phase-matched process \cite{PhysRev.127.1918, 10.5555/1817101}, slowly varying envelope, undepleted and long-wavelength pump approximations, the idler-signal dynamics within the NLPC can be expressed as an optical spin transport equation within a 2D magnetic material \cite{https://doi.org/10.1002/lpor.201300107, PhysRevA.78.063821, Karnieli:18, 10.1063/1.4870695, Sakurai_Napolitano_2020, Karnieli2021, Karnieli2021_FoP} (see Supplemental Material (SM) \cite{supplemental}):
\begin{equation}
\label{eqn:Eq1}
    i \frac{\partial}{\partial z} \Psi(\textbf{r}) = \left[ \frac{\mathbf{p}^{2}_{T}}{2\bar{k}} - \pmb{\sigma} \cdot \mathbf{M}_T(\mathbf{r}) \right] \Psi(\textbf{r}),
\end{equation}
where $\Psi(\textbf{r}) = \frac{1}{\sqrt{N}}\left (\sqrt{\frac{n_{i}}{\omega_{i}}}A_{i}(\textbf{r}), \sqrt{\frac{n_{s}}{\omega_{s}}}A_{s}(\textbf{r})\right)^{T}$ is the effective spinor in the rotating frame, with $N = \frac{n_{i}}{\omega_{i}} \left | A_{i}(\textbf{r}) \right | ^{2} + \frac{n_{s}}{\omega_{s}} \left | A_{s}(\textbf{r}) \right | ^{2}$ a constant of motion due to the Manley-Rowe relations \cite{10.5555/1817101}, such that $||\Psi|| = \sqrt{\Psi^{\dagger} \Psi} = 1$. Here, $A_{j}(\textbf{r})$, $n_{j}$ with $j=i,s$ are the slowly varying envelopes and refractive indices of the idler and signal fields, respectively. $\bar{k}$ is the mean wave number of the idler and signal, $\mathbf{p}_T = -i\nabla_T = -i(\partial_x, \partial_y)$ is the transverse momentum operator in position-space representation, $\pmb{\sigma} = (\sigma_x, \sigma_y, \sigma_z)$ is the Pauli matrix-vector, and $\textbf{r} = (x,y,z)$ is a 3D position vector. $\textbf{M}_T(\textbf{r})$ is the transverse synthetic magnetization texture driving the dynamics between the interacting fields and is defined as $\textbf{M}_T(\textbf{r}) = M_0(\textbf{r}) \hat{\textbf{M}}_T(\phi(\textbf{r})) = M_0(\textbf{r}) \left[ \hat{\textbf{x}} \cos \phi(\textbf{r}) + \hat{\textbf{y}} \sin \phi(\textbf{r}) \right]$. Here, $M_0(\textbf{r}) \propto |\chi^{(2)}(\textbf{r}) A_p(\textbf{r})|$, where $\chi^{(2)}(\textbf{r})$ is the position-dependent second-order susceptibility, $A_p(\textbf{r})$ is the complex envelope of the pump field, and $\phi(\textbf{r}) = \phi_p(\textbf{r}) - \phi_{NLPC}(\textbf{r})$ is the relative phase between the pump envelope and the NLPC modulation.

In this framework, a 1D periodically-poled crystal with a wide Gaussian pump beam generates a uniform transverse magnetization texture, analogous to a ferromagnetic phase (Fig.~\ref{fig:figure1}(a)). To emulate a disordered spin-glass phase, illustrated in Fig.~\ref{fig:figure1}(b), we create a random synthetic magnetization pattern by shaping the pump field into a non-diffracting (ND-) speckle pattern \cite{Schwartz2007, Billy2008, doi:10.1126/science.1209019, Jendrzejewski2012, PhysRevLett.127.180601}, effectively producing propagation-invariant disorder, as required for transverse localization schemes \cite{PhysRevLett.62.47, Schwartz2007, PhysRevLett.100.013906, Segev2013, Mafi:15, 10.1063/1.5142161}. This is depicted on the left of Fig.~\ref{fig:figure1}(b). Alternatively, disorder can be introduced by sampling the crystal's modulation phase, $\phi_{NLPC}(\textbf{r})$, from a uniform distribution within $\left[0, 2\pi\right]_x \times \left[0, 2\pi\right]_y$, while keeping it propagation-invariant by maintaining a constant pump envelope phase along the propagation direction using a wide Gaussian beam, as illustrated on the right of Fig.~\ref{fig:figure1}(b).

While pump shaping can be experimentally implemented with standard optical components and a 1D-modulated NLPC, it is limited by wave diffraction. In contrast, crystal shaping provides greater control over disorder strength and structure by independently tuning magnetization density and orientation, the two degrees of freedom defining vectorial disorder in spin glasses. However, it requires 3D-modulated NLPCs, currently restricted to short crystal lengths \cite{Zhang:21, Imbrock:20, Yu:24}. Consequently, we numerically study pseudospin localization using crystal shaping and experimentally verify our predictions through pump shaping. Additionally, Appendix A demonstrates how current 2D-modulated NLPC technology can induce 1D transverse localization.

\begin{figure}[ht]
\includegraphics[width=\linewidth]{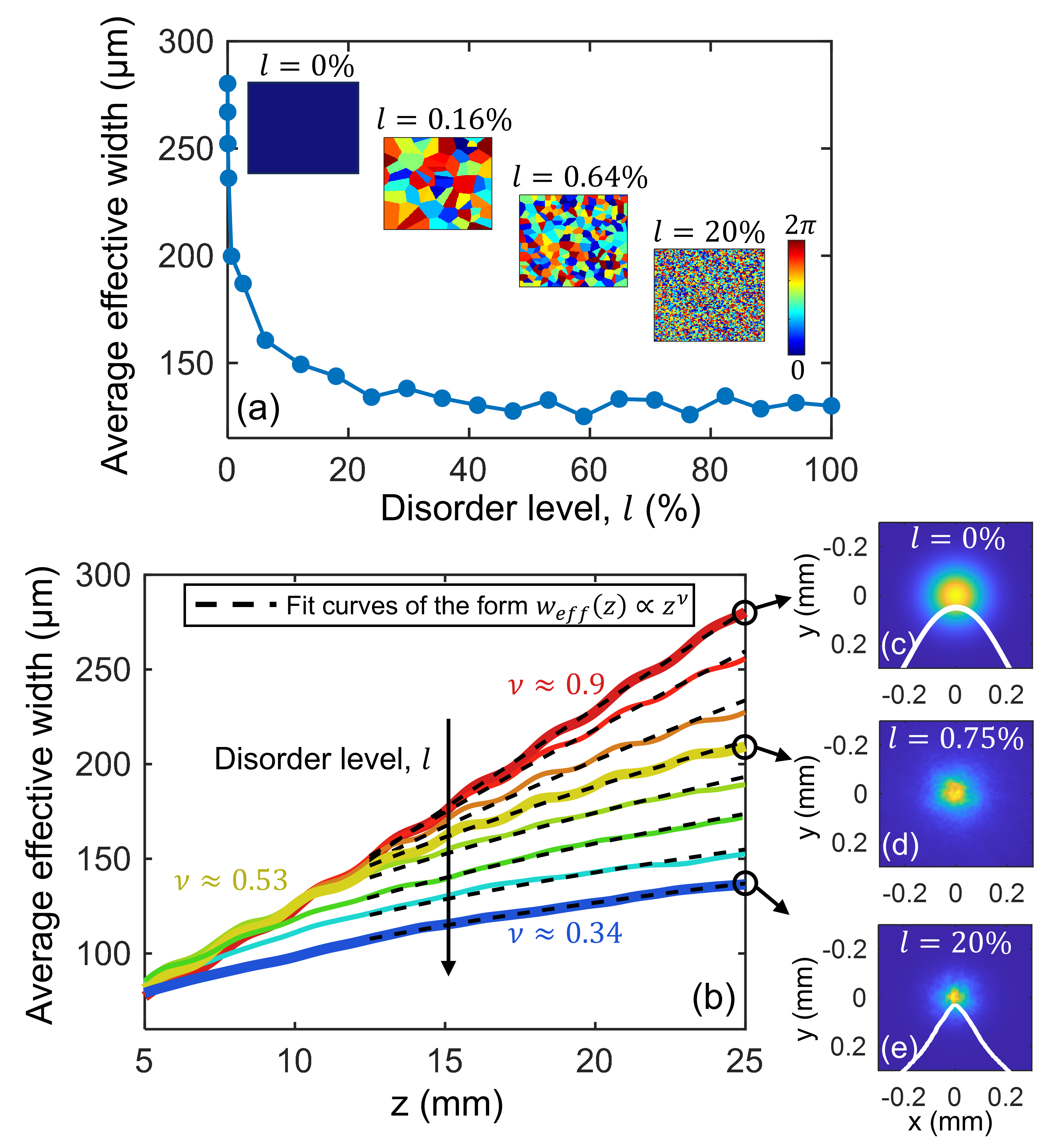}
\caption{\label{fig:figure2}\textbf{Pseudospin transport in a synthetically disordered 3D-modulated NLPC for a fixed pump intensity of 142.5 MW/cm$^2$.} (a) Average effective width at the crystal output versus disorder level. Insets illustrate synthetic magnetization domains with randomly aligned magnetic moments determined by $\phi_{\text{NLPC}}(\mathbf{r}) \in [0, 2\pi]$ for various disorder levels. (b) Average effective width as a function of propagation distance for different disorder levels, with black dashed curves show theoretical fits. (c)-(e) Ensemble-averaged idler-signal output intensity cross-sections for  $l=0\%$, $0.75\%$, and $20\%$, respectively.}
\end{figure}

We simulated the idler-signal pseudospin transport within a 25 mm-long synthetically disordered periodically-poled KTiOPO$_{4}$ (PPKTP) NLPC using the split-step Fourier method \cite{agrawal2013nonlinear}. Given the statistical nature of the phenomenon, our results were obtained by averaging over 70 disorder realizations with identical strength and statistical properties. To quantify the disorder level in the synthetic transverse magnetization, we divided the grid into magnetic domains of varying sizes and random shapes \cite{myref}, with $\phi_{NLPC} (\textbf{r})$ constant within each domain and uniformly sampled from $\left[0, 2\pi\right]$ across domains. The disorder level, defined as $l=100\times \left(N_{d} - 1\right) / \left(N_{p} - 1\right) \left(\% \right)$, is determined by the number of domains, $N_{d}$, ranging from 1 (full order, resembling a ferromagnetic phase), to $N_{p}$ (full disorder, emulating a spin-glass phase), where $N_{p}$ is the total number of pixels in the transverse grid. Intermediate $N_{d}$ values represent a magnetic domain structure \cite{hubert1998magnetic, Sun2021} (see insets of Fig.~\ref{fig:figure2}).

To evaluate the localization strength, we measured the beam confinement during propagation by calculating the average effective beam width, $w_{\text{eff}}(z) = \langle P(z) \rangle^{-1/2}$, where $\langle \cdot \rangle$ denotes averaging over multiple disorder realizations, and $\mathcal{P} \equiv \frac{\iint I(x,y,z)^2 \, dx \, dy}{\left( \iint I(x,y,z) \, dx \, dy \right)^2}$ is the inverse participation ratio \cite{DERAEDT19871, PhysRevLett.62.47, Schwartz2007, PhysRevB.83.184206, Mafi:15, Wang2024}. Here, $I(x,y,z)$ represents the intensity distribution after propagating a distance $z$ through the crystal, with the integration performed over the entire transverse grid.

Figure \ref{fig:figure2}(a) shows the average effective width of the idler-signal beam at the crystal output as a function of disorder level for a fixed pump intensity, revealing that it decreases with increasing disorder, stabilizing above $l=20\%$. Figure \ref{fig:figure2}(b) presents the evolution of the average effective width, $w_{\text{eff}}(z)$, along the propagation axis for different disorder levels. Generally, $w_{\text{eff}}(z)$ follows a power-law behavior with propagation distance, $w_{\text{eff}} \propto z^{\nu}$ \cite{DERAEDT19871, PhysRevLett.62.47, PhysRevLett.96.063904, Mafi:15, Schwartz2007}. In the absence of disorder $\left(l=0\%\right)$, corresponding to an optical ferromagnet, ballistic transport is expected with $w_{\text{eff}}$ increasing linearly with $z$, resulting in $\nu \approx 1$ \cite{DERAEDT19871, PhysRevLett.96.063904, Mafi:15, Schwartz2007}. This is confirmed by the linear fit for $l=0\%$ in Fig.~\ref{fig:figure2}(b) and the wide Gaussian output beam intensity shown in Fig.~\ref{fig:figure2}(c), arising from dominant diffraction effects. Here, the intensity distribution follows $I \propto \exp\left(-2 \frac{\textbf{r}_T^2}{\sigma^2} \right)$, where $\textbf{r}_T = (x,y)$ is the transverse position vector, and $\sigma$ denotes the Gaussian beam width.

With weak disorder, the transport shifts towards diffusion, with $\nu \approx 0.5$ \cite{DERAEDT19871, Schwartz2007}. In our system, at $l=0.75\%$ the best-fit curve yields $\nu \approx 0.53$, with the corresponding ensemble-averaged output intensity shown in Fig.~\ref{fig:figure2}(d). For higher disorder levels, the beam initially undergoes diffusive transport but then localizes after a certain propagation distance. In these cases, $\nu$ decreases to values within $0 < \nu < \frac{1}{2}$, depending on the disorder level \cite{PhysRevLett.62.47, DERAEDT19871, Schwartz2007, PhysRevB.83.184206, PhysRevLett.96.063904, Mafi:15}. As disorder increases, localization becomes stronger, and for disorder levels above $l=20\%$, $\nu$ stabilizes around 0.34. Figure \ref{fig:figure2}(e) shows that the ensemble-averaged output intensity becomes transversely localized, exhibiting an exponential decay profile $\langle I \rangle \propto \exp\left(-2 \frac{|\textbf{r}_T|}{\xi_{\text{loc}}} \right)$, with $|\textbf{r}_T| = \sqrt{x^2 + y^2}$ representing the transverse distance from the beam's center, and $\xi_{\text{loc}}$ is the localization length, defining the beam width where diffusion ceases and localization dominates \cite{myref1}.

The transition from ballistic transport in the 2D optical ferromagnet to localization in the 2D optical spin-glass is further evident when comparing the intensity profiles on a logarithmic scale: a \textit{parabolic} curve for the optical ferromagnet versus a \textit{linear} curve for the optical spin-glass (see insets of Figs.~\ref{fig:figure2}(c) and \ref{fig:figure2}(e), respectively). This highlights that the idler-signal light beam, representing the spin current, undergoes transverse localization due to the induced disorder. Notably, while Fig.~\ref{fig:figure2} demonstrates the localization of the \textit{two-frequency} idler-signal beam under disorder, localization occurs separately for both the idler and signal, with each becoming transversely localized (see SM \cite{supplemental}).

\begin{figure}[ht]
\includegraphics[width=\linewidth]{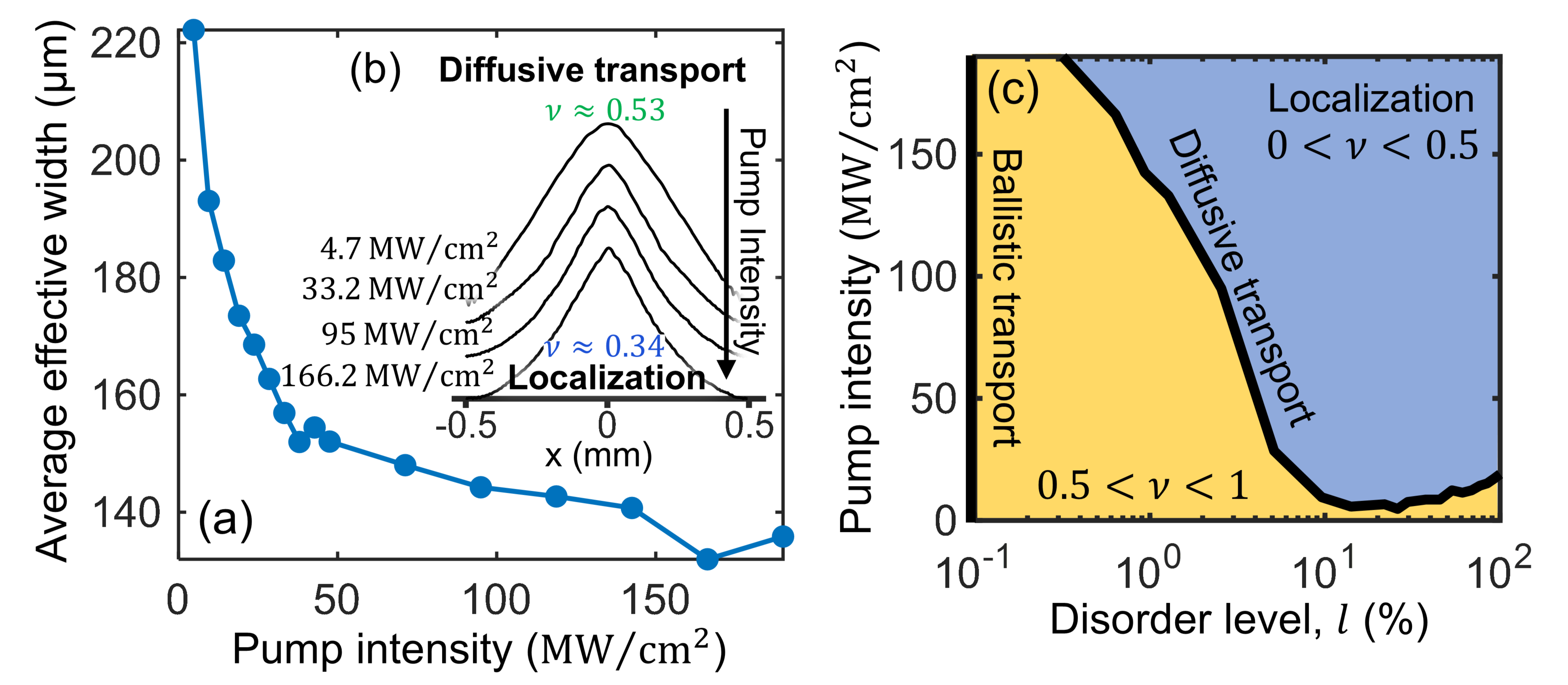}
\caption{\label{fig:figure3}\textbf{Pseudospin localization and nonlinearity.} (a) Average effective width at the crystal output versus pump intensity for a fixed disorder level of $l=26\%$. (b) Logarithmic profiles of the ensemble-averaged idler-signal output intensity at $l=26\%$ for various pump intensities. (c) Phase diagram of pseudospin transport as a function of disorder level and pump intensity: the left black line marks ballistic transport in the absence of disorder ($\nu \approx 1$); the middle line represents diffusive transport ($\nu \approx 0.5$); the yellow shaded region shows the transition from ballistic to diffusive transport; and the blue shaded region indicates a transition to localization, which becomes stronger with increasing nonlinearity and disorder.}
\end{figure}

Our numerical model allows us to isolate the effects of nonlinearity and disorder on this localization phenomenon, enabling us to examine the interplay between the two. Figure ~\ref{fig:figure3}(a) shows that at a fixed disorder level, the average effective width at the crystal output decreases with increasing pump intensity, stabilizing for pump intensities above 100 MW/cm$^2$. This indicates that disorder alone cannot fully explain this localization, and a strong coupling between the idler and signal beams, facilitated by a high pump intensity, is also necessary. If the effect were purely linear and perturbative, localization would occur even with a weak signal field and nearly undepleted idler. However, our findings demonstrate that it is the combination of strong disorder and non-perturbative nonlinear coupling that drives localization. Figure ~\ref{fig:figure3}(b) illustrates this: at $l=26\%$, the logarithmic plot of the ensemble-averaged output intensity for different pump intensities shows that, at low pump intensities, the SFG process occurs without achieving localization, with the intensity profile displaying diffusive transport. However, as the pump intensity increases, the idler-signal beam becomes transversely localized, forming a narrow, linear intensity profile.

The relationship between nonlinearity and localization is summarized in \ref{fig:figure3}(c), showing that stronger nonlinearity (higher pump intensity) not only enhances localization by reducing the localization length but also induces the characteristic exponential decay of localization at lower disorder levels \cite{Schwartz2007, PhysRevLett.100.013906, Segev2013, PhysRevLett.112.193902}. These findings offer new insights into the interplay between nonlinearity and localization, a topic extensively studied in contexts such as $\chi^{(3)}$-nonlinearity \cite{PhysRevLett.69.1807, Schwartz2007, PhysRevLett.100.013906, PhysRevLett.112.193902}, and $\chi^{(2)}$-nonlinearity of second harmonic generation \cite{AGRANOVICH1988378, Yoo:89}. From a spintronics perspective, our results suggest that in spin-glass phases, where the magnetization texture introduces a disordered vectorial potential, the magnetization density (analogous to nonlinearity strength in our optical system) plays a central role in spin current localization, alongside the randomness in magnetic moment alignment.

Another interesting aspect of this localization involves the energy exchange between the idler and signal fields, which can be modeled as a two-level system coupled by the nonlinear coupling \cite{PhysRevA.78.063821, https://doi.org/10.1002/lpor.201300107, Karnieli2021_FoP}. Under phase-matching, this coupling leads to Rabi oscillations, where energy oscillates between the idler and signal \cite{PhysRevA.78.063821, Karnieli2021_FoP}. However, in the optical disordered spin-glass, where pseudospin localization occurs, our simulations reveal a pseudospin decoherence effect, manifesting as decaying Rabi oscillations. This effect is further discussed in Appendix B.

To experimentally verify our predictions, we employed the pump shaping approach and created an all-optical disordered transverse synthetic magnetization pattern using an ND-speckled pump. The experimental setup, detailed in SM \cite{supplemental}, involved an SFG process within a 1D-modulated PPKTP crystal, between a pulsed pump laser at 1064.5 nm and a continuous-wave idler laser at 1550 nm. The ND-speckled pump beam was generated using an axicon lens and rotating diffuser, producing multiple disorder realizations with identical properties \cite{Schwartz2007, PhysRevLett.127.180601}. To corroborate our results, we simulated the interaction using the pump shaping approach with the same experimental parameters.

\begin{figure}[ht]
\includegraphics[width=\linewidth]{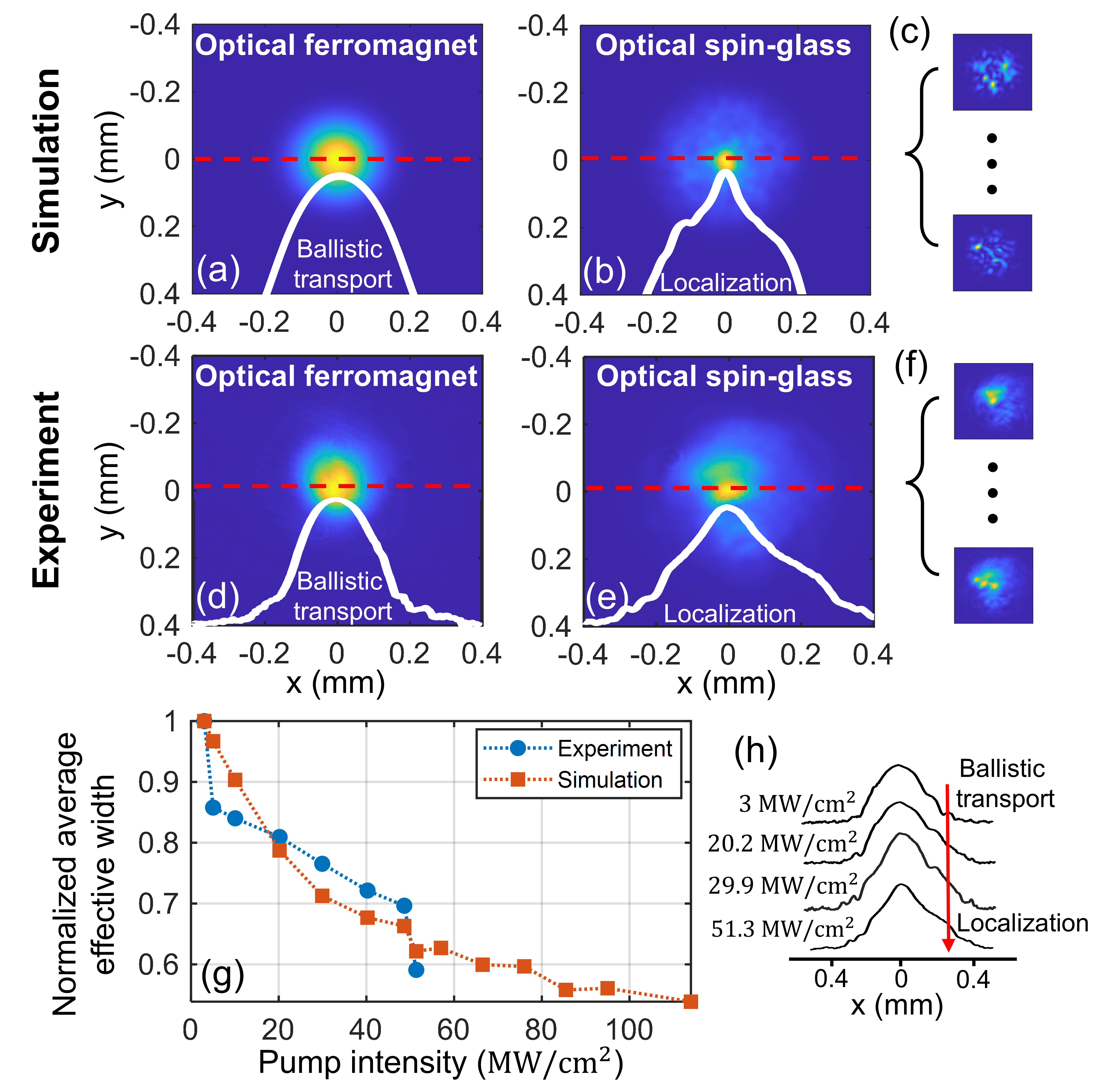}
\caption{\label{fig:figure4}\textbf{All-optical pseudospin localization by pump shaping.} (a)-(d) 2D cross-sections of the signal intensity at the crystal output for an optical ferromagnet (a, d) and spin-glass (b, e)as obtained in the simulation and experiment, respectively. Insets show the corresponding logarithmic profiles (white) along the horizontal red dashed lines. Panels (b) and (e) display ensembled-averaged signal output intensity, with two specific realizations shown in panels (c) and (f), for the simulation and experiment, respectively. The pump intensity in (a)-(f) is $\sim$51.4 MW/cm$^2$. (g) Normalized average effective width of the signal at the crystal output versus pump intensity. (h) Measured ensemble-averaged signal output intensity logarithmic profiles for different pump intensities.}
\end{figure}

When employing a Gaussian pump beam, effectively creating an optical ferromagnet, we observed a broad Gaussian signal beam at the crystal output, as shown in Figs.~\ref{fig:figure4}(a) and ~\ref{fig:figure4}(d) for the simulation and experiment, respectively. This reflects dominant diffraction effects, corresponding to ballistic transport of spin current in a ferromagnetic phase. Conversely, when employing an ND-speckled pump beam, effectively creating an optical disordered spin-glass, the signal diffraction is suppressed. As predicted, the ensemble-averaged signal output intensity becomes transversely localized with exponentially decaying tails, as shown in Figs.~\ref{fig:figure4}(b) and ~\ref{fig:figure4}(e) for the simulation and experiment, respectively. Importantly, the localization signature is evident from the linear profiles on a logarithmic scale for the optical spin-glass (insets of Figs.~\ref{fig:figure4}(b) and ~\ref{fig:figure4}(e)), compared to the parabolic profiles in the optical ferromagnet (insets of Figs.~\ref{fig:figure4}(a) and ~\ref{fig:figure4}(d)) \cite{footnote4}.

Figure ~\ref{fig:figure4}(g) shows the normalized effective width of the signal at the crystal output, $w_{\text{eff}}(L)$, at increasing pump intensities. As the intensity approaches the maximum intensity available in our experimental setup (51.4 MW/cm$^2$), $w_{\text{eff}}(L)$ decreases, with our simulation predicting further narrowing at higher intensities. Furthermore, examining the logarithmic ensemble-averaged signal output intensity profiles for various pump intensities, as depicted in Fig.~\ref{fig:figure4}(h), reveals that at low intensities, localization is not yet achieved, and the profile is non-exponential. As the pump intensity increases, the signal beam becomes transversely localized, with the logarithmic profile transitioning to a clear linear shape, and the localization length decreases. These findings confirm our numerical predictions, demonstrating that both high pump intensity and disorder are necessary for idler-signal pseudospin localization.

To summarize, we have numerically and experimentally demonstrated a new, controllable transverse localization of light phenomenon in a nonlinear optical system analogous to a disordered spin-glass phase, enhancing our understanding of localization in disordered vectorial potentials. This may shed light on phenomena such as ferromagnetic-metal to spin-glass-insulator phase transitions \cite{Villain1979, GOODENOUGH2002297, PhysRevLett.99.086401, PhysRevB.101.205106}, similar to metal-insulator transitions in disordered scalar potentials \cite{PhysRev.109.1492}. Our controllable system opens avenues for investigating other complex magnetization textures, including spin-glass phases with structured and correlated disorder \cite{RevModPhys.58.801, Nishimori_2022, Krempaský2023, Mézard2024}, and spin-ice phases \cite{Skjærvø2020, Lendinez_2020, 10.1063/5.0046083}. Additionally, expanding synthetic disorder to all three dimensions of the NLPC could emulate temporal modulation in spintronic systems \cite{Karnieli2021}, allowing exploration beyond equilibrium dynamics.

Extending our framework to the depleted pump regime, where idler/signal back-action perturbs the synthetic magnetization \cite{Karnieli2021, Karnieli2021_FoP}, could reveal further details about the interplay between nonlinearity and localization. This may also provide insights into static and dynamic spin-transfer torque in spin-glass \cite{PhysRevB.96.100402}. Additionally, our approach can be applied to other nonlinear processes, such as difference frequency generation \cite{Mondal:22}, $\chi^{(3)}$-nonlinearity \cite{PhysRevLett.124.153902, Li:21, Bashan:24}, and other two-level systems, including polarization states of light \cite{https://doi.org/10.1002/lpor.202301030, Liu2023, Liu:24}, and coupled optical waveguides \cite{https://doi.org/10.1002/lpor.202301055}, potentially uncovering additional light localization phenomena.

Finally, our framework can be expanded into the quantum domain, where spin currents could be implemented using quantum light. This opens the possibility of exploring quantum spin phenomena in vectorial potentials with two-frequency idler-signal entangled photons \cite{Kobayashi2016, Karnieli:18, Kues2019, PhysRevLett.124.143601, Lu:23}, including studies of quantum correlations in quantum random walks with disorder \cite{PhysRevA.48.1687, PhysRevLett.105.163905, PhysRevLett.102.253904, doi:10.1126/science.1193515}.\\

\textit{Acknowledgments}—S.I. acknowledges support from the Colton Scholarship Foundation and the Weinstein Institute for signal processing. A.K. acknowledges support from the Urbanek-Chodorow postdoctoral fellowship of Stanford University and the Zuckerman STEM leadership postdoctoral program. S.T. acknowledges support from the Yad Hanadiv Foundation through the Rothschild fellowship and the Helen Diller Quantum Center postdoctoral fellowship. Both A.K. and S.T. acknowledge support from the VATAT-Quantum fellowship and the Viterbi fellowship of the Technion—Israel Institute of Technology. This work was funded by the Israel Science Foundation, grants 969/22 and 3117/23.

\bibliography{main_text_inc_figures}

\begin{thebibliography}{107}%
\makeatletter
\providecommand \@ifxundefined [1]{%
 \@ifx{#1\undefined}
}%
\providecommand \@ifnum [1]{%
 \ifnum #1\expandafter \@firstoftwo
 \else \expandafter \@secondoftwo
 \fi
}%
\providecommand \@ifx [1]{%
 \ifx #1\expandafter \@firstoftwo
 \else \expandafter \@secondoftwo
 \fi
}%
\providecommand \natexlab [1]{#1}%
\providecommand \enquote  [1]{``#1''}%
\providecommand \bibnamefont  [1]{#1}%
\providecommand \bibfnamefont [1]{#1}%
\providecommand \citenamefont [1]{#1}%
\providecommand \href@noop [0]{\@secondoftwo}%
\providecommand \href [0]{\begingroup \@sanitize@url \@href}%
\providecommand \@href[1]{\@@startlink{#1}\@@href}%
\providecommand \@@href[1]{\endgroup#1\@@endlink}%
\providecommand \@sanitize@url [0]{\catcode `\\12\catcode `\$12\catcode `\&12\catcode `\#12\catcode `\^12\catcode `\_12\catcode `\%12\relax}%
\providecommand \@@startlink[1]{}%
\providecommand \@@endlink[0]{}%
\providecommand \url  [0]{\begingroup\@sanitize@url \@url }%
\providecommand \@url [1]{\endgroup\@href {#1}{\urlprefix }}%
\providecommand \urlprefix  [0]{URL }%
\providecommand \Eprint [0]{\href }%
\providecommand \doibase [0]{https://doi.org/}%
\providecommand \selectlanguage [0]{\@gobble}%
\providecommand \bibinfo  [0]{\@secondoftwo}%
\providecommand \bibfield  [0]{\@secondoftwo}%
\providecommand \translation [1]{[#1]}%
\providecommand \BibitemOpen [0]{}%
\providecommand \bibitemStop [0]{}%
\providecommand \bibitemNoStop [0]{.\EOS\space}%
\providecommand \EOS [0]{\spacefactor3000\relax}%
\providecommand \BibitemShut  [1]{\csname bibitem#1\endcsname}%
\let\auto@bib@innerbib\@empty
\bibitem [{\citenamefont {Wolf}\ \emph {et~al.}(2001)\citenamefont {Wolf}, \citenamefont {Awschalom}, \citenamefont {Buhrman}, \citenamefont {Daughton}, \citenamefont {von Molnár}, \citenamefont {Roukes}, \citenamefont {Chtchelkanova},\ and\ \citenamefont {Treger}}]{doi:10.1126/science.1065389}%
  \BibitemOpen
  \bibfield  {author} {\bibinfo {author} {\bibfnamefont {S.~A.}\ \bibnamefont {Wolf}}, \bibinfo {author} {\bibfnamefont {D.~D.}\ \bibnamefont {Awschalom}}, \bibinfo {author} {\bibfnamefont {R.~A.}\ \bibnamefont {Buhrman}}, \bibinfo {author} {\bibfnamefont {J.~M.}\ \bibnamefont {Daughton}}, \bibinfo {author} {\bibfnamefont {S.}~\bibnamefont {von Molnár}}, \bibinfo {author} {\bibfnamefont {M.~L.}\ \bibnamefont {Roukes}}, \bibinfo {author} {\bibfnamefont {A.~Y.}\ \bibnamefont {Chtchelkanova}},\ and\ \bibinfo {author} {\bibfnamefont {D.~M.}\ \bibnamefont {Treger}},\ }\bibfield  {title} {\bibinfo {title} {Spintronics: A spin-based electronics vision for the future},\ }\href {https://doi.org/10.1126/science.1065389} {\bibfield  {journal} {\bibinfo  {journal} {Science}\ }\textbf {\bibinfo {volume} {294}},\ \bibinfo {pages} {1488} (\bibinfo {year} {2001})}\BibitemShut {NoStop}%
\bibitem [{\citenamefont {Edwards}\ and\ \citenamefont {Anderson}(1975)}]{S_F_Edwards_1975}%
  \BibitemOpen
  \bibfield  {author} {\bibinfo {author} {\bibfnamefont {S.~F.}\ \bibnamefont {Edwards}}\ and\ \bibinfo {author} {\bibfnamefont {P.~W.}\ \bibnamefont {Anderson}},\ }\bibfield  {title} {\bibinfo {title} {Theory of spin glasses},\ }\href {https://doi.org/10.1088/0305-4608/5/5/017} {\bibfield  {journal} {\bibinfo  {journal} {Journal of Physics F: Metal Physics}\ }\textbf {\bibinfo {volume} {5}},\ \bibinfo {pages} {965} (\bibinfo {year} {1975})}\BibitemShut {NoStop}%
\bibitem [{\citenamefont {M\'ezard}\ \emph {et~al.}(1984)\citenamefont {M\'ezard}, \citenamefont {Parisi}, \citenamefont {Sourlas}, \citenamefont {Toulouse},\ and\ \citenamefont {Virasoro}}]{PhysRevLett.52.1156}%
  \BibitemOpen
  \bibfield  {author} {\bibinfo {author} {\bibfnamefont {M.}~\bibnamefont {M\'ezard}}, \bibinfo {author} {\bibfnamefont {G.}~\bibnamefont {Parisi}}, \bibinfo {author} {\bibfnamefont {N.}~\bibnamefont {Sourlas}}, \bibinfo {author} {\bibfnamefont {G.}~\bibnamefont {Toulouse}},\ and\ \bibinfo {author} {\bibfnamefont {M.}~\bibnamefont {Virasoro}},\ }\bibfield  {title} {\bibinfo {title} {Nature of the spin-glass phase},\ }\href {https://doi.org/10.1103/PhysRevLett.52.1156} {\bibfield  {journal} {\bibinfo  {journal} {Phys. Rev. Lett.}\ }\textbf {\bibinfo {volume} {52}},\ \bibinfo {pages} {1156} (\bibinfo {year} {1984})}\BibitemShut {NoStop}%
\bibitem [{\citenamefont {Mezard}\ \emph {et~al.}(1987)\citenamefont {Mezard}, \citenamefont {Parisi},\ and\ \citenamefont {Virasoro}}]{mezard1987spin}%
  \BibitemOpen
  \bibfield  {author} {\bibinfo {author} {\bibfnamefont {M.}~\bibnamefont {Mezard}}, \bibinfo {author} {\bibfnamefont {G.}~\bibnamefont {Parisi}},\ and\ \bibinfo {author} {\bibfnamefont {M.}~\bibnamefont {Virasoro}},\ }\href {https://books.google.co.il/books?id=DwY8DQAAQBAJ} {\emph {\bibinfo {title} {Spin Glass Theory And Beyond: An Introduction To The Replica Method And Its Applications}}},\ World Scientific Lecture Notes In Physics\ (\bibinfo  {publisher} {World Scientific Publishing Company},\ \bibinfo {year} {1987})\BibitemShut {NoStop}%
\bibitem [{\citenamefont {Weissman}(1993)}]{RevModPhys.65.829}%
  \BibitemOpen
  \bibfield  {author} {\bibinfo {author} {\bibfnamefont {M.~B.}\ \bibnamefont {Weissman}},\ }\bibfield  {title} {\bibinfo {title} {What is a spin glass? a glimpse via mesoscopic noise},\ }\href {https://doi.org/10.1103/RevModPhys.65.829} {\bibfield  {journal} {\bibinfo  {journal} {Rev. Mod. Phys.}\ }\textbf {\bibinfo {volume} {65}},\ \bibinfo {pages} {829} (\bibinfo {year} {1993})}\BibitemShut {NoStop}%
\bibitem [{\citenamefont {Parisi}(2006)}]{doi:10.1073/pnas.0601120103}%
  \BibitemOpen
  \bibfield  {author} {\bibinfo {author} {\bibfnamefont {G.}~\bibnamefont {Parisi}},\ }\bibfield  {title} {\bibinfo {title} {Spin glasses and fragile glasses: Statics, dynamics, and complexity},\ }\href {https://doi.org/10.1073/pnas.0601120103} {\bibfield  {journal} {\bibinfo  {journal} {Proceedings of the National Academy of Sciences}\ }\textbf {\bibinfo {volume} {103}},\ \bibinfo {pages} {7948} (\bibinfo {year} {2006})}\BibitemShut {NoStop}%
\bibitem [{\citenamefont {Parisi}(1983)}]{PhysRevLett.50.1946}%
  \BibitemOpen
  \bibfield  {author} {\bibinfo {author} {\bibfnamefont {G.}~\bibnamefont {Parisi}},\ }\bibfield  {title} {\bibinfo {title} {Order parameter for spin-glasses},\ }\href {https://doi.org/10.1103/PhysRevLett.50.1946} {\bibfield  {journal} {\bibinfo  {journal} {Phys. Rev. Lett.}\ }\textbf {\bibinfo {volume} {50}},\ \bibinfo {pages} {1946} (\bibinfo {year} {1983})}\BibitemShut {NoStop}%
\bibitem [{\citenamefont {Venturelli}\ \emph {et~al.}(2015)\citenamefont {Venturelli}, \citenamefont {Mandr\`a}, \citenamefont {Knysh}, \citenamefont {O'Gorman}, \citenamefont {Biswas},\ and\ \citenamefont {Smelyanskiy}}]{PhysRevX.5.031040}%
  \BibitemOpen
  \bibfield  {author} {\bibinfo {author} {\bibfnamefont {D.}~\bibnamefont {Venturelli}}, \bibinfo {author} {\bibfnamefont {S.}~\bibnamefont {Mandr\`a}}, \bibinfo {author} {\bibfnamefont {S.}~\bibnamefont {Knysh}}, \bibinfo {author} {\bibfnamefont {B.}~\bibnamefont {O'Gorman}}, \bibinfo {author} {\bibfnamefont {R.}~\bibnamefont {Biswas}},\ and\ \bibinfo {author} {\bibfnamefont {V.}~\bibnamefont {Smelyanskiy}},\ }\bibfield  {title} {\bibinfo {title} {Quantum optimization of fully connected spin glasses},\ }\href {https://doi.org/10.1103/PhysRevX.5.031040} {\bibfield  {journal} {\bibinfo  {journal} {Phys. Rev. X}\ }\textbf {\bibinfo {volume} {5}},\ \bibinfo {pages} {031040} (\bibinfo {year} {2015})}\BibitemShut {NoStop}%
\bibitem [{\citenamefont {Fan}\ \emph {et~al.}(2023)\citenamefont {Fan}, \citenamefont {Shen}, \citenamefont {Nussinov}, \citenamefont {Liu}, \citenamefont {Sun},\ and\ \citenamefont {Liu}}]{Fan2023}%
  \BibitemOpen
  \bibfield  {author} {\bibinfo {author} {\bibfnamefont {C.}~\bibnamefont {Fan}}, \bibinfo {author} {\bibfnamefont {M.}~\bibnamefont {Shen}}, \bibinfo {author} {\bibfnamefont {Z.}~\bibnamefont {Nussinov}}, \bibinfo {author} {\bibfnamefont {Z.}~\bibnamefont {Liu}}, \bibinfo {author} {\bibfnamefont {Y.}~\bibnamefont {Sun}},\ and\ \bibinfo {author} {\bibfnamefont {Y.-Y.}\ \bibnamefont {Liu}},\ }\bibfield  {title} {\bibinfo {title} {Searching for spin glass ground states through deep reinforcement learning},\ }\href {https://doi.org/10.1038/s41467-023-36363-w} {\bibfield  {journal} {\bibinfo  {journal} {Nature Communications}\ }\textbf {\bibinfo {volume} {14}},\ \bibinfo {pages} {725} (\bibinfo {year} {2023})}\BibitemShut {NoStop}%
\bibitem [{Nob(2021)}]{NobelPrize2021}%
  \BibitemOpen
  \href {\url{https://www.nobelprize.org/prizes/physics/2021/summary/}} {\bibinfo {title} {The nobel prize in physics 2021 \url{https://www.nobelprize.org/prizes/physics/2021/summary/}}} (\bibinfo {year} {2021})\BibitemShut {NoStop}%
\bibitem [{\citenamefont {Binder}\ and\ \citenamefont {Young}(1986)}]{RevModPhys.58.801}%
  \BibitemOpen
  \bibfield  {author} {\bibinfo {author} {\bibfnamefont {K.}~\bibnamefont {Binder}}\ and\ \bibinfo {author} {\bibfnamefont {A.~P.}\ \bibnamefont {Young}},\ }\bibfield  {title} {\bibinfo {title} {Spin glasses: Experimental facts, theoretical concepts, and open questions},\ }\href {https://doi.org/10.1103/RevModPhys.58.801} {\bibfield  {journal} {\bibinfo  {journal} {Rev. Mod. Phys.}\ }\textbf {\bibinfo {volume} {58}},\ \bibinfo {pages} {801} (\bibinfo {year} {1986})}\BibitemShut {NoStop}%
\bibitem [{\citenamefont {Goodenough}\ \emph {et~al.}(2002)\citenamefont {Goodenough}, \citenamefont {Dass},\ and\ \citenamefont {Zhou}}]{GOODENOUGH2002297}%
  \BibitemOpen
  \bibfield  {author} {\bibinfo {author} {\bibfnamefont {J.~B.}\ \bibnamefont {Goodenough}}, \bibinfo {author} {\bibfnamefont {R.~I.}\ \bibnamefont {Dass}},\ and\ \bibinfo {author} {\bibfnamefont {J.}~\bibnamefont {Zhou}},\ }\bibfield  {title} {\bibinfo {title} {Spin-glass to ferromagnet transition in {LaMn$_{1-x}$Sc$_x$O$_3$}},\ }\href {https://doi.org/https://doi.org/10.1016/S1293-2558(01)01255-9} {\bibfield  {journal} {\bibinfo  {journal} {Solid State Sciences}\ }\textbf {\bibinfo {volume} {4}},\ \bibinfo {pages} {297} (\bibinfo {year} {2002})}\BibitemShut {NoStop}%
\bibitem [{\citenamefont {Hanasaki}\ \emph {et~al.}(2007)\citenamefont {Hanasaki}, \citenamefont {Watanabe}, \citenamefont {Ohtsuka}, \citenamefont {K\'ezsm\'arki}, \citenamefont {Iguchi}, \citenamefont {Miyasaka},\ and\ \citenamefont {Tokura}}]{PhysRevLett.99.086401}%
  \BibitemOpen
  \bibfield  {author} {\bibinfo {author} {\bibfnamefont {N.}~\bibnamefont {Hanasaki}}, \bibinfo {author} {\bibfnamefont {K.}~\bibnamefont {Watanabe}}, \bibinfo {author} {\bibfnamefont {T.}~\bibnamefont {Ohtsuka}}, \bibinfo {author} {\bibfnamefont {I.}~\bibnamefont {K\'ezsm\'arki}}, \bibinfo {author} {\bibfnamefont {S.}~\bibnamefont {Iguchi}}, \bibinfo {author} {\bibfnamefont {S.}~\bibnamefont {Miyasaka}},\ and\ \bibinfo {author} {\bibfnamefont {Y.}~\bibnamefont {Tokura}},\ }\bibfield  {title} {\bibinfo {title} {Nature of the transition between a ferromagnetic metal and a spin-glass insulator in pyrochlore molybdates},\ }\href {https://doi.org/10.1103/PhysRevLett.99.086401} {\bibfield  {journal} {\bibinfo  {journal} {Phys. Rev. Lett.}\ }\textbf {\bibinfo {volume} {99}},\ \bibinfo {pages} {086401} (\bibinfo {year} {2007})}\BibitemShut {NoStop}%
\bibitem [{\citenamefont {Appelbaum}\ \emph {et~al.}(2007)\citenamefont {Appelbaum}, \citenamefont {Huang},\ and\ \citenamefont {Monsma}}]{Appelbaum2007}%
  \BibitemOpen
  \bibfield  {author} {\bibinfo {author} {\bibfnamefont {I.}~\bibnamefont {Appelbaum}}, \bibinfo {author} {\bibfnamefont {B.}~\bibnamefont {Huang}},\ and\ \bibinfo {author} {\bibfnamefont {D.~J.}\ \bibnamefont {Monsma}},\ }\bibfield  {title} {\bibinfo {title} {Electronic measurement and control of spin transport in silicon},\ }\href {https://doi.org/10.1038/nature05803} {\bibfield  {journal} {\bibinfo  {journal} {Nature}\ }\textbf {\bibinfo {volume} {447}},\ \bibinfo {pages} {295} (\bibinfo {year} {2007})}\BibitemShut {NoStop}%
\bibitem [{\citenamefont {Liu}\ and\ \citenamefont {Vignale}(2011)}]{PhysRevLett.106.247203}%
  \BibitemOpen
  \bibfield  {author} {\bibinfo {author} {\bibfnamefont {T.}~\bibnamefont {Liu}}\ and\ \bibinfo {author} {\bibfnamefont {G.}~\bibnamefont {Vignale}},\ }\bibfield  {title} {\bibinfo {title} {Electric control of spin currents and spin-wave logic},\ }\href {https://doi.org/10.1103/PhysRevLett.106.247203} {\bibfield  {journal} {\bibinfo  {journal} {Phys. Rev. Lett.}\ }\textbf {\bibinfo {volume} {106}},\ \bibinfo {pages} {247203} (\bibinfo {year} {2011})}\BibitemShut {NoStop}%
\bibitem [{\citenamefont {Shi}\ \emph {et~al.}(2021)\citenamefont {Shi}, \citenamefont {Nisoli},\ and\ \citenamefont {Chern}}]{10.1063/5.0046083}%
  \BibitemOpen
  \bibfield  {author} {\bibinfo {author} {\bibfnamefont {Y.}~\bibnamefont {Shi}}, \bibinfo {author} {\bibfnamefont {C.}~\bibnamefont {Nisoli}},\ and\ \bibinfo {author} {\bibfnamefont {G.-W.}\ \bibnamefont {Chern}},\ }\bibfield  {title} {\bibinfo {title} {{Ice, glass, and solid phases in artificial spin systems with quenched disorder}},\ }\href {https://doi.org/10.1063/5.0046083} {\bibfield  {journal} {\bibinfo  {journal} {Applied Physics Letters}\ }\textbf {\bibinfo {volume} {118}},\ \bibinfo {pages} {122407} (\bibinfo {year} {2021})}\BibitemShut {NoStop}%
\bibitem [{\citenamefont {Han}\ \emph {et~al.}(2020)\citenamefont {Han}, \citenamefont {Zhang}, \citenamefont {Bi}, \citenamefont {Fan}, \citenamefont {Safi}, \citenamefont {Xiang}, \citenamefont {Finley}, \citenamefont {Fu}, \citenamefont {Cheng},\ and\ \citenamefont {Liu}}]{Han2020}%
  \BibitemOpen
  \bibfield  {author} {\bibinfo {author} {\bibfnamefont {J.}~\bibnamefont {Han}}, \bibinfo {author} {\bibfnamefont {P.}~\bibnamefont {Zhang}}, \bibinfo {author} {\bibfnamefont {Z.}~\bibnamefont {Bi}}, \bibinfo {author} {\bibfnamefont {Y.}~\bibnamefont {Fan}}, \bibinfo {author} {\bibfnamefont {T.~S.}\ \bibnamefont {Safi}}, \bibinfo {author} {\bibfnamefont {J.}~\bibnamefont {Xiang}}, \bibinfo {author} {\bibfnamefont {J.}~\bibnamefont {Finley}}, \bibinfo {author} {\bibfnamefont {L.}~\bibnamefont {Fu}}, \bibinfo {author} {\bibfnamefont {R.}~\bibnamefont {Cheng}},\ and\ \bibinfo {author} {\bibfnamefont {L.}~\bibnamefont {Liu}},\ }\bibfield  {title} {\bibinfo {title} {Birefringence-like spin transport via linearly polarized antiferromagnetic magnons},\ }\href {https://doi.org/10.1038/s41565-020-0703-8} {\bibfield  {journal} {\bibinfo  {journal} {Nature Nanotechnology}\ }\textbf {\bibinfo {volume} {15}},\ \bibinfo {pages} {563} (\bibinfo {year} {2020})}\BibitemShut {NoStop}%
\bibitem [{\citenamefont {Chen}\ \emph {et~al.}(2022)\citenamefont {Chen}, \citenamefont {Bai}, \citenamefont {Ji}, \citenamefont {Zhou}, \citenamefont {Liao}, \citenamefont {You}, \citenamefont {Zhu}, \citenamefont {Wang}, \citenamefont {Han}, \citenamefont {Liu}, \citenamefont {Li}, \citenamefont {Han}, \citenamefont {Yin}, \citenamefont {Kou}, \citenamefont {Pan},\ and\ \citenamefont {Song}}]{Chen2022}%
  \BibitemOpen
  \bibfield  {author} {\bibinfo {author} {\bibfnamefont {X.}~\bibnamefont {Chen}}, \bibinfo {author} {\bibfnamefont {H.}~\bibnamefont {Bai}}, \bibinfo {author} {\bibfnamefont {Y.}~\bibnamefont {Ji}}, \bibinfo {author} {\bibfnamefont {Y.}~\bibnamefont {Zhou}}, \bibinfo {author} {\bibfnamefont {L.}~\bibnamefont {Liao}}, \bibinfo {author} {\bibfnamefont {Y.}~\bibnamefont {You}}, \bibinfo {author} {\bibfnamefont {W.}~\bibnamefont {Zhu}}, \bibinfo {author} {\bibfnamefont {Q.}~\bibnamefont {Wang}}, \bibinfo {author} {\bibfnamefont {L.}~\bibnamefont {Han}}, \bibinfo {author} {\bibfnamefont {X.}~\bibnamefont {Liu}}, \bibinfo {author} {\bibfnamefont {A.}~\bibnamefont {Li}}, \bibinfo {author} {\bibfnamefont {X.}~\bibnamefont {Han}}, \bibinfo {author} {\bibfnamefont {J.}~\bibnamefont {Yin}}, \bibinfo {author} {\bibfnamefont {X.}~\bibnamefont {Kou}}, \bibinfo {author} {\bibfnamefont {F.}~\bibnamefont {Pan}},\ and\ \bibinfo {author} {\bibfnamefont {C.}~\bibnamefont {Song}},\ }\bibfield  {title} {\bibinfo {title} {Control
  of spin current and antiferromagnetic moments via topological surface state},\ }\href {https://doi.org/10.1038/s41928-022-00825-8} {\bibfield  {journal} {\bibinfo  {journal} {Nature Electronics}\ }\textbf {\bibinfo {volume} {5}},\ \bibinfo {pages} {574} (\bibinfo {year} {2022})}\BibitemShut {NoStop}%
\bibitem [{\citenamefont {Albada}\ and\ \citenamefont {Lagendijk}(1985)}]{PhysRevLett.55.2692}%
  \BibitemOpen
  \bibfield  {author} {\bibinfo {author} {\bibfnamefont {M.~P.~V.}\ \bibnamefont {Albada}}\ and\ \bibinfo {author} {\bibfnamefont {A.}~\bibnamefont {Lagendijk}},\ }\bibfield  {title} {\bibinfo {title} {Observation of weak localization of light in a random medium},\ }\href {https://doi.org/10.1103/PhysRevLett.55.2692} {\bibfield  {journal} {\bibinfo  {journal} {Phys. Rev. Lett.}\ }\textbf {\bibinfo {volume} {55}},\ \bibinfo {pages} {2692} (\bibinfo {year} {1985})}\BibitemShut {NoStop}%
\bibitem [{\citenamefont {Wolf}\ and\ \citenamefont {Maret}(1985)}]{PhysRevLett.55.2696}%
  \BibitemOpen
  \bibfield  {author} {\bibinfo {author} {\bibfnamefont {P.-E.}\ \bibnamefont {Wolf}}\ and\ \bibinfo {author} {\bibfnamefont {G.}~\bibnamefont {Maret}},\ }\bibfield  {title} {\bibinfo {title} {Weak localization and coherent backscattering of photons in disordered media},\ }\href {https://doi.org/10.1103/PhysRevLett.55.2696} {\bibfield  {journal} {\bibinfo  {journal} {Phys. Rev. Lett.}\ }\textbf {\bibinfo {volume} {55}},\ \bibinfo {pages} {2696} (\bibinfo {year} {1985})}\BibitemShut {NoStop}%
\bibitem [{\citenamefont {Anderson}(1958)}]{PhysRev.109.1492}%
  \BibitemOpen
  \bibfield  {author} {\bibinfo {author} {\bibfnamefont {P.~W.}\ \bibnamefont {Anderson}},\ }\bibfield  {title} {\bibinfo {title} {Absence of diffusion in certain random lattices},\ }\href {https://doi.org/10.1103/PhysRev.109.1492} {\bibfield  {journal} {\bibinfo  {journal} {Phys. Rev.}\ }\textbf {\bibinfo {volume} {109}},\ \bibinfo {pages} {1492} (\bibinfo {year} {1958})}\BibitemShut {NoStop}%
\bibitem [{\citenamefont {Lagendijk}\ \emph {et~al.}(2009)\citenamefont {Lagendijk}, \citenamefont {Tiggelen},\ and\ \citenamefont {Wiersma}}]{10.1063/1.3206091}%
  \BibitemOpen
  \bibfield  {author} {\bibinfo {author} {\bibfnamefont {A.}~\bibnamefont {Lagendijk}}, \bibinfo {author} {\bibfnamefont {B.~v.}\ \bibnamefont {Tiggelen}},\ and\ \bibinfo {author} {\bibfnamefont {D.~S.}\ \bibnamefont {Wiersma}},\ }\bibfield  {title} {\bibinfo {title} {{Fifty years of Anderson localization}},\ }\href {https://doi.org/10.1063/1.3206091} {\bibfield  {journal} {\bibinfo  {journal} {Physics Today}\ }\textbf {\bibinfo {volume} {62}},\ \bibinfo {pages} {24} (\bibinfo {year} {2009})}\BibitemShut {NoStop}%
\bibitem [{\citenamefont {Segev}\ \emph {et~al.}(2013)\citenamefont {Segev}, \citenamefont {Silberberg},\ and\ \citenamefont {Christodoulides}}]{Segev2013}%
  \BibitemOpen
  \bibfield  {author} {\bibinfo {author} {\bibfnamefont {M.}~\bibnamefont {Segev}}, \bibinfo {author} {\bibfnamefont {Y.}~\bibnamefont {Silberberg}},\ and\ \bibinfo {author} {\bibfnamefont {D.~N.}\ \bibnamefont {Christodoulides}},\ }\bibfield  {title} {\bibinfo {title} {Anderson localization of light},\ }\href {https://doi.org/10.1038/nphoton.2013.30} {\bibfield  {journal} {\bibinfo  {journal} {Nature Photonics}\ }\textbf {\bibinfo {volume} {7}},\ \bibinfo {pages} {197} (\bibinfo {year} {2013})}\BibitemShut {NoStop}%
\bibitem [{\citenamefont {Agranovich}\ and\ \citenamefont {Kravtsov}(1988)}]{AGRANOVICH1988378}%
  \BibitemOpen
  \bibfield  {author} {\bibinfo {author} {\bibfnamefont {V.}~\bibnamefont {Agranovich}}\ and\ \bibinfo {author} {\bibfnamefont {V.}~\bibnamefont {Kravtsov}},\ }\bibfield  {title} {\bibinfo {title} {Effects of weak localization of photons in nonlinear optics: Second harmonic generation},\ }\href {https://doi.org/https://doi.org/10.1016/0375-9601(88)90792-X} {\bibfield  {journal} {\bibinfo  {journal} {Physics Letters A}\ }\textbf {\bibinfo {volume} {131}},\ \bibinfo {pages} {378} (\bibinfo {year} {1988})}\BibitemShut {NoStop}%
\bibitem [{\citenamefont {de~Boer}\ \emph {et~al.}(1993)\citenamefont {de~Boer}, \citenamefont {Lagendijk}, \citenamefont {Sprik},\ and\ \citenamefont {Feng}}]{PhysRevLett.71.3947}%
  \BibitemOpen
  \bibfield  {author} {\bibinfo {author} {\bibfnamefont {J.~F.}\ \bibnamefont {de~Boer}}, \bibinfo {author} {\bibfnamefont {A.}~\bibnamefont {Lagendijk}}, \bibinfo {author} {\bibfnamefont {R.}~\bibnamefont {Sprik}},\ and\ \bibinfo {author} {\bibfnamefont {S.}~\bibnamefont {Feng}},\ }\bibfield  {title} {\bibinfo {title} {Transmission and reflection correlations of second harmonic waves in nonlinear random media},\ }\href {https://doi.org/10.1103/PhysRevLett.71.3947} {\bibfield  {journal} {\bibinfo  {journal} {Phys. Rev. Lett.}\ }\textbf {\bibinfo {volume} {71}},\ \bibinfo {pages} {3947} (\bibinfo {year} {1993})}\BibitemShut {NoStop}%
\bibitem [{\citenamefont {Wiersma}\ \emph {et~al.}(1997)\citenamefont {Wiersma}, \citenamefont {Bartolini}, \citenamefont {Lagendijk},\ and\ \citenamefont {Righini}}]{Wiersma1997}%
  \BibitemOpen
  \bibfield  {author} {\bibinfo {author} {\bibfnamefont {D.~S.}\ \bibnamefont {Wiersma}}, \bibinfo {author} {\bibfnamefont {P.}~\bibnamefont {Bartolini}}, \bibinfo {author} {\bibfnamefont {A.}~\bibnamefont {Lagendijk}},\ and\ \bibinfo {author} {\bibfnamefont {R.}~\bibnamefont {Righini}},\ }\bibfield  {title} {\bibinfo {title} {Localization of light in a disordered medium},\ }\href {https://doi.org/10.1038/37757} {\bibfield  {journal} {\bibinfo  {journal} {Nature}\ }\textbf {\bibinfo {volume} {390}},\ \bibinfo {pages} {671} (\bibinfo {year} {1997})}\BibitemShut {NoStop}%
\bibitem [{\citenamefont {Berry}\ and\ \citenamefont {Klein}(1997)}]{M_V_Berry_1997}%
  \BibitemOpen
  \bibfield  {author} {\bibinfo {author} {\bibfnamefont {M.~V.}\ \bibnamefont {Berry}}\ and\ \bibinfo {author} {\bibfnamefont {S.}~\bibnamefont {Klein}},\ }\bibfield  {title} {\bibinfo {title} {Transparent mirrors: rays, waves and localization},\ }\href {https://doi.org/10.1088/0143-0807/18/3/017} {\bibfield  {journal} {\bibinfo  {journal} {European Journal of Physics}\ }\textbf {\bibinfo {volume} {18}},\ \bibinfo {pages} {222} (\bibinfo {year} {1997})}\BibitemShut {NoStop}%
\bibitem [{\citenamefont {Chabanov}\ \emph {et~al.}(2000)\citenamefont {Chabanov}, \citenamefont {Stoytchev},\ and\ \citenamefont {Genack}}]{Chabanov2000}%
  \BibitemOpen
  \bibfield  {author} {\bibinfo {author} {\bibfnamefont {A.~A.}\ \bibnamefont {Chabanov}}, \bibinfo {author} {\bibfnamefont {M.}~\bibnamefont {Stoytchev}},\ and\ \bibinfo {author} {\bibfnamefont {A.~Z.}\ \bibnamefont {Genack}},\ }\bibfield  {title} {\bibinfo {title} {Statistical signatures of photon localization},\ }\href {https://doi.org/10.1038/35009055} {\bibfield  {journal} {\bibinfo  {journal} {Nature}\ }\textbf {\bibinfo {volume} {404}},\ \bibinfo {pages} {850} (\bibinfo {year} {2000})}\BibitemShut {NoStop}%
\bibitem [{\citenamefont {Faez}\ \emph {et~al.}(2009)\citenamefont {Faez}, \citenamefont {Johnson}, \citenamefont {Mazurenko},\ and\ \citenamefont {Lagendijk}}]{Faez:09}%
  \BibitemOpen
  \bibfield  {author} {\bibinfo {author} {\bibfnamefont {S.}~\bibnamefont {Faez}}, \bibinfo {author} {\bibfnamefont {P.~M.}\ \bibnamefont {Johnson}}, \bibinfo {author} {\bibfnamefont {D.~A.}\ \bibnamefont {Mazurenko}},\ and\ \bibinfo {author} {\bibfnamefont {A.}~\bibnamefont {Lagendijk}},\ }\bibfield  {title} {\bibinfo {title} {Experimental observation of second-harmonic generation and diffusion inside random media},\ }\href {https://doi.org/10.1364/JOSAB.26.000235} {\bibfield  {journal} {\bibinfo  {journal} {J. Opt. Soc. Am. B}\ }\textbf {\bibinfo {volume} {26}},\ \bibinfo {pages} {235} (\bibinfo {year} {2009})}\BibitemShut {NoStop}%
\bibitem [{\citenamefont {St\"orzer}\ \emph {et~al.}(2006)\citenamefont {St\"orzer}, \citenamefont {Gross}, \citenamefont {Aegerter},\ and\ \citenamefont {Maret}}]{PhysRevLett.96.063904}%
  \BibitemOpen
  \bibfield  {author} {\bibinfo {author} {\bibfnamefont {M.}~\bibnamefont {St\"orzer}}, \bibinfo {author} {\bibfnamefont {P.}~\bibnamefont {Gross}}, \bibinfo {author} {\bibfnamefont {C.~M.}\ \bibnamefont {Aegerter}},\ and\ \bibinfo {author} {\bibfnamefont {G.}~\bibnamefont {Maret}},\ }\bibfield  {title} {\bibinfo {title} {Observation of the critical regime near anderson localization of light},\ }\href {https://doi.org/10.1103/PhysRevLett.96.063904} {\bibfield  {journal} {\bibinfo  {journal} {Phys. Rev. Lett.}\ }\textbf {\bibinfo {volume} {96}},\ \bibinfo {pages} {063904} (\bibinfo {year} {2006})}\BibitemShut {NoStop}%
\bibitem [{\citenamefont {Samanta}\ \emph {et~al.}(2023)\citenamefont {Samanta}, \citenamefont {Pierrat}, \citenamefont {Carminati},\ and\ \citenamefont {Mujumdar}}]{PhysRevA.108.053501}%
  \BibitemOpen
  \bibfield  {author} {\bibinfo {author} {\bibfnamefont {R.}~\bibnamefont {Samanta}}, \bibinfo {author} {\bibfnamefont {R.}~\bibnamefont {Pierrat}}, \bibinfo {author} {\bibfnamefont {R.}~\bibnamefont {Carminati}},\ and\ \bibinfo {author} {\bibfnamefont {S.}~\bibnamefont {Mujumdar}},\ }\bibfield  {title} {\bibinfo {title} {Photon diffusion in space and time in a second-order-nonlinear disordered medium},\ }\href {https://doi.org/10.1103/PhysRevA.108.053501} {\bibfield  {journal} {\bibinfo  {journal} {Phys. Rev. A}\ }\textbf {\bibinfo {volume} {108}},\ \bibinfo {pages} {053501} (\bibinfo {year} {2023})}\BibitemShut {NoStop}%
\bibitem [{\citenamefont {Yamilov}\ \emph {et~al.}(2023)\citenamefont {Yamilov}, \citenamefont {Skipetrov}, \citenamefont {Hughes}, \citenamefont {Minkov}, \citenamefont {Yu},\ and\ \citenamefont {Cao}}]{Yamilov2023}%
  \BibitemOpen
  \bibfield  {author} {\bibinfo {author} {\bibfnamefont {A.}~\bibnamefont {Yamilov}}, \bibinfo {author} {\bibfnamefont {S.~E.}\ \bibnamefont {Skipetrov}}, \bibinfo {author} {\bibfnamefont {T.~W.}\ \bibnamefont {Hughes}}, \bibinfo {author} {\bibfnamefont {M.}~\bibnamefont {Minkov}}, \bibinfo {author} {\bibfnamefont {Z.}~\bibnamefont {Yu}},\ and\ \bibinfo {author} {\bibfnamefont {H.}~\bibnamefont {Cao}},\ }\bibfield  {title} {\bibinfo {title} {Anderson localization of electromagnetic waves in three dimensions},\ }\href {https://doi.org/10.1038/s41567-023-02091-7} {\bibfield  {journal} {\bibinfo  {journal} {Nature Physics}\ }\textbf {\bibinfo {volume} {19}},\ \bibinfo {pages} {1308} (\bibinfo {year} {2023})}\BibitemShut {NoStop}%
\bibitem [{\citenamefont {Schwartz}\ \emph {et~al.}(2007)\citenamefont {Schwartz}, \citenamefont {Bartal}, \citenamefont {Fishman},\ and\ \citenamefont {Segev}}]{Schwartz2007}%
  \BibitemOpen
  \bibfield  {author} {\bibinfo {author} {\bibfnamefont {T.}~\bibnamefont {Schwartz}}, \bibinfo {author} {\bibfnamefont {G.}~\bibnamefont {Bartal}}, \bibinfo {author} {\bibfnamefont {S.}~\bibnamefont {Fishman}},\ and\ \bibinfo {author} {\bibfnamefont {M.}~\bibnamefont {Segev}},\ }\bibfield  {title} {\bibinfo {title} {Transport and anderson localization in disordered two-dimensional photonic lattices},\ }\href {https://doi.org/10.1038/nature05623} {\bibfield  {journal} {\bibinfo  {journal} {Nature}\ }\textbf {\bibinfo {volume} {446}},\ \bibinfo {pages} {52} (\bibinfo {year} {2007})}\BibitemShut {NoStop}%
\bibitem [{\citenamefont {Lahini}\ \emph {et~al.}(2008)\citenamefont {Lahini}, \citenamefont {Avidan}, \citenamefont {Pozzi}, \citenamefont {Sorel}, \citenamefont {Morandotti}, \citenamefont {Christodoulides},\ and\ \citenamefont {Silberberg}}]{PhysRevLett.100.013906}%
  \BibitemOpen
  \bibfield  {author} {\bibinfo {author} {\bibfnamefont {Y.}~\bibnamefont {Lahini}}, \bibinfo {author} {\bibfnamefont {A.}~\bibnamefont {Avidan}}, \bibinfo {author} {\bibfnamefont {F.}~\bibnamefont {Pozzi}}, \bibinfo {author} {\bibfnamefont {M.}~\bibnamefont {Sorel}}, \bibinfo {author} {\bibfnamefont {R.}~\bibnamefont {Morandotti}}, \bibinfo {author} {\bibfnamefont {D.~N.}\ \bibnamefont {Christodoulides}},\ and\ \bibinfo {author} {\bibfnamefont {Y.}~\bibnamefont {Silberberg}},\ }\bibfield  {title} {\bibinfo {title} {Anderson localization and nonlinearity in one-dimensional disordered photonic lattices},\ }\href {https://doi.org/10.1103/PhysRevLett.100.013906} {\bibfield  {journal} {\bibinfo  {journal} {Phys. Rev. Lett.}\ }\textbf {\bibinfo {volume} {100}},\ \bibinfo {pages} {013906} (\bibinfo {year} {2008})}\BibitemShut {NoStop}%
\bibitem [{\citenamefont {Karbasi}\ \emph {et~al.}(2012)\citenamefont {Karbasi}, \citenamefont {Mirr}, \citenamefont {Yarandi}, \citenamefont {Frazier}, \citenamefont {Koch},\ and\ \citenamefont {Mafi}}]{Karbasi:12}%
  \BibitemOpen
  \bibfield  {author} {\bibinfo {author} {\bibfnamefont {S.}~\bibnamefont {Karbasi}}, \bibinfo {author} {\bibfnamefont {C.~R.}\ \bibnamefont {Mirr}}, \bibinfo {author} {\bibfnamefont {P.~G.}\ \bibnamefont {Yarandi}}, \bibinfo {author} {\bibfnamefont {R.~J.}\ \bibnamefont {Frazier}}, \bibinfo {author} {\bibfnamefont {K.~W.}\ \bibnamefont {Koch}},\ and\ \bibinfo {author} {\bibfnamefont {A.}~\bibnamefont {Mafi}},\ }\bibfield  {title} {\bibinfo {title} {Observation of transverse anderson localization in an optical fiber},\ }\href {https://doi.org/10.1364/OL.37.002304} {\bibfield  {journal} {\bibinfo  {journal} {Opt. Lett.}\ }\textbf {\bibinfo {volume} {37}},\ \bibinfo {pages} {2304} (\bibinfo {year} {2012})}\BibitemShut {NoStop}%
\bibitem [{\citenamefont {Boguslawski}\ \emph {et~al.}(2013)\citenamefont {Boguslawski}, \citenamefont {Brake}, \citenamefont {Armijo}, \citenamefont {Diebel}, \citenamefont {Rose},\ and\ \citenamefont {Denz}}]{Boguslawski:13}%
  \BibitemOpen
  \bibfield  {author} {\bibinfo {author} {\bibfnamefont {M.}~\bibnamefont {Boguslawski}}, \bibinfo {author} {\bibfnamefont {S.}~\bibnamefont {Brake}}, \bibinfo {author} {\bibfnamefont {J.}~\bibnamefont {Armijo}}, \bibinfo {author} {\bibfnamefont {F.}~\bibnamefont {Diebel}}, \bibinfo {author} {\bibfnamefont {P.}~\bibnamefont {Rose}},\ and\ \bibinfo {author} {\bibfnamefont {C.}~\bibnamefont {Denz}},\ }\bibfield  {title} {\bibinfo {title} {Analysis of transverse anderson localization in refractive index structures with customized random potential},\ }\href {https://doi.org/10.1364/OE.21.031713} {\bibfield  {journal} {\bibinfo  {journal} {Opt. Express}\ }\textbf {\bibinfo {volume} {21}},\ \bibinfo {pages} {31713} (\bibinfo {year} {2013})}\BibitemShut {NoStop}%
\bibitem [{\citenamefont {Leonetti}\ \emph {et~al.}(2014)\citenamefont {Leonetti}, \citenamefont {Karbasi}, \citenamefont {Mafi},\ and\ \citenamefont {Conti}}]{PhysRevLett.112.193902}%
  \BibitemOpen
  \bibfield  {author} {\bibinfo {author} {\bibfnamefont {M.}~\bibnamefont {Leonetti}}, \bibinfo {author} {\bibfnamefont {S.}~\bibnamefont {Karbasi}}, \bibinfo {author} {\bibfnamefont {A.}~\bibnamefont {Mafi}},\ and\ \bibinfo {author} {\bibfnamefont {C.}~\bibnamefont {Conti}},\ }\bibfield  {title} {\bibinfo {title} {Observation of migrating transverse anderson localizations of light in nonlocal media},\ }\href {https://doi.org/10.1103/PhysRevLett.112.193902} {\bibfield  {journal} {\bibinfo  {journal} {Phys. Rev. Lett.}\ }\textbf {\bibinfo {volume} {112}},\ \bibinfo {pages} {193902} (\bibinfo {year} {2014})}\BibitemShut {NoStop}%
\bibitem [{\citenamefont {Dikopoltsev}\ \emph {et~al.}(2019)\citenamefont {Dikopoltsev}, \citenamefont {Herzig~Sheinfux},\ and\ \citenamefont {Segev}}]{PhysRevB.100.140202}%
  \BibitemOpen
  \bibfield  {author} {\bibinfo {author} {\bibfnamefont {A.}~\bibnamefont {Dikopoltsev}}, \bibinfo {author} {\bibfnamefont {H.}~\bibnamefont {Herzig~Sheinfux}},\ and\ \bibinfo {author} {\bibfnamefont {M.}~\bibnamefont {Segev}},\ }\bibfield  {title} {\bibinfo {title} {Localization by virtual transitions in correlated disorder},\ }\href {https://doi.org/10.1103/PhysRevB.100.140202} {\bibfield  {journal} {\bibinfo  {journal} {Phys. Rev. B}\ }\textbf {\bibinfo {volume} {100}},\ \bibinfo {pages} {140202} (\bibinfo {year} {2019})}\BibitemShut {NoStop}%
\bibitem [{\citenamefont {Gianfrate}\ \emph {et~al.}(2020)\citenamefont {Gianfrate}, \citenamefont {Dominici}, \citenamefont {Ballarini}, \citenamefont {Sanvitto},\ and\ \citenamefont {Leonetti}}]{10.1063/1.5142161}%
  \BibitemOpen
  \bibfield  {author} {\bibinfo {author} {\bibfnamefont {A.}~\bibnamefont {Gianfrate}}, \bibinfo {author} {\bibfnamefont {L.}~\bibnamefont {Dominici}}, \bibinfo {author} {\bibfnamefont {D.}~\bibnamefont {Ballarini}}, \bibinfo {author} {\bibfnamefont {D.}~\bibnamefont {Sanvitto}},\ and\ \bibinfo {author} {\bibfnamefont {M.}~\bibnamefont {Leonetti}},\ }\bibfield  {title} {\bibinfo {title} {{Transverse localization of light in laser written designed disorder}},\ }\href {https://doi.org/10.1063/1.5142161} {\bibfield  {journal} {\bibinfo  {journal} {Applied Physics Letters}\ }\textbf {\bibinfo {volume} {116}},\ \bibinfo {pages} {071101} (\bibinfo {year} {2020})}\BibitemShut {NoStop}%
\bibitem [{\citenamefont {Dikopoltsev}\ \emph {et~al.}(2022)\citenamefont {Dikopoltsev}, \citenamefont {Weidemann}, \citenamefont {Kremer}, \citenamefont {Steinfurth}, \citenamefont {Sheinfux}, \citenamefont {Szameit},\ and\ \citenamefont {Segev}}]{doi:10.1126/sciadv.abn7769}%
  \BibitemOpen
  \bibfield  {author} {\bibinfo {author} {\bibfnamefont {A.}~\bibnamefont {Dikopoltsev}}, \bibinfo {author} {\bibfnamefont {S.}~\bibnamefont {Weidemann}}, \bibinfo {author} {\bibfnamefont {M.}~\bibnamefont {Kremer}}, \bibinfo {author} {\bibfnamefont {A.}~\bibnamefont {Steinfurth}}, \bibinfo {author} {\bibfnamefont {H.~H.}\ \bibnamefont {Sheinfux}}, \bibinfo {author} {\bibfnamefont {A.}~\bibnamefont {Szameit}},\ and\ \bibinfo {author} {\bibfnamefont {M.}~\bibnamefont {Segev}},\ }\bibfield  {title} {\bibinfo {title} {Observation of anderson localization beyond the spectrum of the disorder},\ }\href {https://doi.org/10.1126/sciadv.abn7769} {\bibfield  {journal} {\bibinfo  {journal} {Science Advances}\ }\textbf {\bibinfo {volume} {8}},\ \bibinfo {pages} {eabn7769} (\bibinfo {year} {2022})}\BibitemShut {NoStop}%
\bibitem [{\citenamefont {McKenna}\ \emph {et~al.}(1992)\citenamefont {McKenna}, \citenamefont {Stanley},\ and\ \citenamefont {Maynard}}]{PhysRevLett.69.1807}%
  \BibitemOpen
  \bibfield  {author} {\bibinfo {author} {\bibfnamefont {M.~J.}\ \bibnamefont {McKenna}}, \bibinfo {author} {\bibfnamefont {R.~L.}\ \bibnamefont {Stanley}},\ and\ \bibinfo {author} {\bibfnamefont {J.~D.}\ \bibnamefont {Maynard}},\ }\bibfield  {title} {\bibinfo {title} {Effects of nonlinearity on anderson localization},\ }\href {https://doi.org/10.1103/PhysRevLett.69.1807} {\bibfield  {journal} {\bibinfo  {journal} {Phys. Rev. Lett.}\ }\textbf {\bibinfo {volume} {69}},\ \bibinfo {pages} {1807} (\bibinfo {year} {1992})}\BibitemShut {NoStop}%
\bibitem [{\citenamefont {Fishman}\ \emph {et~al.}(2012)\citenamefont {Fishman}, \citenamefont {Krivolapov},\ and\ \citenamefont {Soffer}}]{Fishman_2012}%
  \BibitemOpen
  \bibfield  {author} {\bibinfo {author} {\bibfnamefont {S.}~\bibnamefont {Fishman}}, \bibinfo {author} {\bibfnamefont {Y.}~\bibnamefont {Krivolapov}},\ and\ \bibinfo {author} {\bibfnamefont {A.}~\bibnamefont {Soffer}},\ }\bibfield  {title} {\bibinfo {title} {The nonlinear schrödinger equation with a random potential: results and puzzles},\ }\href {https://doi.org/10.1088/0951-7715/25/4/R53} {\bibfield  {journal} {\bibinfo  {journal} {Nonlinearity}\ }\textbf {\bibinfo {volume} {25}},\ \bibinfo {pages} {R53} (\bibinfo {year} {2012})}\BibitemShut {NoStop}%
\bibitem [{\citenamefont {Mafi}(2015)}]{Mafi:15}%
  \BibitemOpen
  \bibfield  {author} {\bibinfo {author} {\bibfnamefont {A.}~\bibnamefont {Mafi}},\ }\bibfield  {title} {\bibinfo {title} {Transverse anderson localization of light: a tutorial},\ }\href {https://doi.org/10.1364/AOP.7.000459} {\bibfield  {journal} {\bibinfo  {journal} {Adv. Opt. Photon.}\ }\textbf {\bibinfo {volume} {7}},\ \bibinfo {pages} {459} (\bibinfo {year} {2015})}\BibitemShut {NoStop}%
\bibitem [{\citenamefont {Sharabi}\ \emph {et~al.}(2018)\citenamefont {Sharabi}, \citenamefont {Herzig~Sheinfux}, \citenamefont {Sagi}, \citenamefont {Eisenstein},\ and\ \citenamefont {Segev}}]{PhysRevLett.121.233901}%
  \BibitemOpen
  \bibfield  {author} {\bibinfo {author} {\bibfnamefont {Y.}~\bibnamefont {Sharabi}}, \bibinfo {author} {\bibfnamefont {H.}~\bibnamefont {Herzig~Sheinfux}}, \bibinfo {author} {\bibfnamefont {Y.}~\bibnamefont {Sagi}}, \bibinfo {author} {\bibfnamefont {G.}~\bibnamefont {Eisenstein}},\ and\ \bibinfo {author} {\bibfnamefont {M.}~\bibnamefont {Segev}},\ }\bibfield  {title} {\bibinfo {title} {Self-induced diffusion in disordered nonlinear photonic media},\ }\href {https://doi.org/10.1103/PhysRevLett.121.233901} {\bibfield  {journal} {\bibinfo  {journal} {Phys. Rev. Lett.}\ }\textbf {\bibinfo {volume} {121}},\ \bibinfo {pages} {233901} (\bibinfo {year} {2018})}\BibitemShut {NoStop}%
\bibitem [{\citenamefont {Suchowski}\ \emph {et~al.}(2014)\citenamefont {Suchowski}, \citenamefont {Porat},\ and\ \citenamefont {Arie}}]{https://doi.org/10.1002/lpor.201300107}%
  \BibitemOpen
  \bibfield  {author} {\bibinfo {author} {\bibfnamefont {H.}~\bibnamefont {Suchowski}}, \bibinfo {author} {\bibfnamefont {G.}~\bibnamefont {Porat}},\ and\ \bibinfo {author} {\bibfnamefont {A.}~\bibnamefont {Arie}},\ }\bibfield  {title} {\bibinfo {title} {Adiabatic processes in frequency conversion},\ }\href {https://doi.org/https://doi.org/10.1002/lpor.201300107} {\bibfield  {journal} {\bibinfo  {journal} {Laser \& Photonics Reviews}\ }\textbf {\bibinfo {volume} {8}},\ \bibinfo {pages} {333} (\bibinfo {year} {2014})}\BibitemShut {NoStop}%
\bibitem [{\citenamefont {Westerberg}\ \emph {et~al.}(2016)\citenamefont {Westerberg}, \citenamefont {Maitland}, \citenamefont {Faccio}, \citenamefont {Wilson}, \citenamefont {\"Ohberg},\ and\ \citenamefont {Wright}}]{PhysRevA.94.023805}%
  \BibitemOpen
  \bibfield  {author} {\bibinfo {author} {\bibfnamefont {N.}~\bibnamefont {Westerberg}}, \bibinfo {author} {\bibfnamefont {C.}~\bibnamefont {Maitland}}, \bibinfo {author} {\bibfnamefont {D.}~\bibnamefont {Faccio}}, \bibinfo {author} {\bibfnamefont {K.}~\bibnamefont {Wilson}}, \bibinfo {author} {\bibfnamefont {P.}~\bibnamefont {\"Ohberg}},\ and\ \bibinfo {author} {\bibfnamefont {E.~M.}\ \bibnamefont {Wright}},\ }\bibfield  {title} {\bibinfo {title} {Synthetic magnetism for photon fluids},\ }\href {https://doi.org/10.1103/PhysRevA.94.023805} {\bibfield  {journal} {\bibinfo  {journal} {Phys. Rev. A}\ }\textbf {\bibinfo {volume} {94}},\ \bibinfo {pages} {023805} (\bibinfo {year} {2016})}\BibitemShut {NoStop}%
\bibitem [{\citenamefont {Karnieli}\ and\ \citenamefont {Arie}(2018{\natexlab{a}})}]{PhysRevLett.120.053901}%
  \BibitemOpen
  \bibfield  {author} {\bibinfo {author} {\bibfnamefont {A.}~\bibnamefont {Karnieli}}\ and\ \bibinfo {author} {\bibfnamefont {A.}~\bibnamefont {Arie}},\ }\bibfield  {title} {\bibinfo {title} {All-optical stern-gerlach effect},\ }\href {https://doi.org/10.1103/PhysRevLett.120.053901} {\bibfield  {journal} {\bibinfo  {journal} {Phys. Rev. Lett.}\ }\textbf {\bibinfo {volume} {120}},\ \bibinfo {pages} {053901} (\bibinfo {year} {2018}{\natexlab{a}})}\BibitemShut {NoStop}%
\bibitem [{\citenamefont {Karnieli}\ and\ \citenamefont {Arie}(2018{\natexlab{b}})}]{Karnieli:18}%
  \BibitemOpen
  \bibfield  {author} {\bibinfo {author} {\bibfnamefont {A.}~\bibnamefont {Karnieli}}\ and\ \bibinfo {author} {\bibfnamefont {A.}~\bibnamefont {Arie}},\ }\bibfield  {title} {\bibinfo {title} {Frequency domain stern-gerlach effect for photonic qubits and qutrits},\ }\href {https://doi.org/10.1364/OPTICA.5.001297} {\bibfield  {journal} {\bibinfo  {journal} {Optica}\ }\textbf {\bibinfo {volume} {5}},\ \bibinfo {pages} {1297} (\bibinfo {year} {2018}{\natexlab{b}})}\BibitemShut {NoStop}%
\bibitem [{\citenamefont {Karnieli}\ \emph {et~al.}(2021{\natexlab{a}})\citenamefont {Karnieli}, \citenamefont {Tsesses}, \citenamefont {Bartal},\ and\ \citenamefont {Arie}}]{Karnieli2021}%
  \BibitemOpen
  \bibfield  {author} {\bibinfo {author} {\bibfnamefont {A.}~\bibnamefont {Karnieli}}, \bibinfo {author} {\bibfnamefont {S.}~\bibnamefont {Tsesses}}, \bibinfo {author} {\bibfnamefont {G.}~\bibnamefont {Bartal}},\ and\ \bibinfo {author} {\bibfnamefont {A.}~\bibnamefont {Arie}},\ }\bibfield  {title} {\bibinfo {title} {Emulating spin transport with nonlinear optics, from high-order skyrmions to the topological hall effect},\ }\href {https://doi.org/10.1038/s41467-021-21250-z} {\bibfield  {journal} {\bibinfo  {journal} {Nature Communications}\ }\textbf {\bibinfo {volume} {12}},\ \bibinfo {pages} {1092} (\bibinfo {year} {2021}{\natexlab{a}})}\BibitemShut {NoStop}%
\bibitem [{\citenamefont {Karnieli}\ \emph {et~al.}(2021{\natexlab{b}})\citenamefont {Karnieli}, \citenamefont {Li},\ and\ \citenamefont {Arie}}]{Karnieli2021_FoP}%
  \BibitemOpen
  \bibfield  {author} {\bibinfo {author} {\bibfnamefont {A.}~\bibnamefont {Karnieli}}, \bibinfo {author} {\bibfnamefont {Y.}~\bibnamefont {Li}},\ and\ \bibinfo {author} {\bibfnamefont {A.}~\bibnamefont {Arie}},\ }\bibfield  {title} {\bibinfo {title} {The geometric phase in nonlinear frequency conversion},\ }\href {https://doi.org/10.1007/s11467-021-1102-9} {\bibfield  {journal} {\bibinfo  {journal} {Frontiers of Physics}\ }\textbf {\bibinfo {volume} {17}},\ \bibinfo {pages} {12301} (\bibinfo {year} {2021}{\natexlab{b}})}\BibitemShut {NoStop}%
\bibitem [{\citenamefont {Yesharim}\ \emph {et~al.}(2022)\citenamefont {Yesharim}, \citenamefont {Karnieli}, \citenamefont {Jackel}, \citenamefont {Di~Domenico}, \citenamefont {Trajtenberg-Mills},\ and\ \citenamefont {Arie}}]{Yesharim2022}%
  \BibitemOpen
  \bibfield  {author} {\bibinfo {author} {\bibfnamefont {O.}~\bibnamefont {Yesharim}}, \bibinfo {author} {\bibfnamefont {A.}~\bibnamefont {Karnieli}}, \bibinfo {author} {\bibfnamefont {S.}~\bibnamefont {Jackel}}, \bibinfo {author} {\bibfnamefont {G.}~\bibnamefont {Di~Domenico}}, \bibinfo {author} {\bibfnamefont {S.}~\bibnamefont {Trajtenberg-Mills}},\ and\ \bibinfo {author} {\bibfnamefont {A.}~\bibnamefont {Arie}},\ }\bibfield  {title} {\bibinfo {title} {Observation of the all-optical stern--gerlach effect in nonlinear optics},\ }\href {https://doi.org/10.1038/s41566-022-01035-6} {\bibfield  {journal} {\bibinfo  {journal} {Nature Photonics}\ }\textbf {\bibinfo {volume} {16}},\ \bibinfo {pages} {582} (\bibinfo {year} {2022})}\BibitemShut {NoStop}%
\bibitem [{\citenamefont {Izhak}\ \emph {et~al.}(2024)\citenamefont {Izhak}, \citenamefont {Karnieli}, \citenamefont {Yesharim}, \citenamefont {Tsesses},\ and\ \citenamefont {Arie}}]{Izhak:24}%
  \BibitemOpen
  \bibfield  {author} {\bibinfo {author} {\bibfnamefont {S.}~\bibnamefont {Izhak}}, \bibinfo {author} {\bibfnamefont {A.}~\bibnamefont {Karnieli}}, \bibinfo {author} {\bibfnamefont {O.}~\bibnamefont {Yesharim}}, \bibinfo {author} {\bibfnamefont {S.}~\bibnamefont {Tsesses}},\ and\ \bibinfo {author} {\bibfnamefont {A.}~\bibnamefont {Arie}},\ }\bibfield  {title} {\bibinfo {title} {All-optical spin valve effect in nonlinear optics},\ }\href {https://doi.org/10.1364/OL.517794} {\bibfield  {journal} {\bibinfo  {journal} {Opt. Lett.}\ }\textbf {\bibinfo {volume} {49}},\ \bibinfo {pages} {1025} (\bibinfo {year} {2024})}\BibitemShut {NoStop}%
\bibitem [{\citenamefont {Liu}\ \emph {et~al.}(2023)\citenamefont {Liu}, \citenamefont {Zhang}, \citenamefont {Zhang}, \citenamefont {Hu}, \citenamefont {Li}, \citenamefont {Chen},\ and\ \citenamefont {Fu}}]{Liu2023}%
  \BibitemOpen
  \bibfield  {author} {\bibinfo {author} {\bibfnamefont {G.}~\bibnamefont {Liu}}, \bibinfo {author} {\bibfnamefont {X.}~\bibnamefont {Zhang}}, \bibinfo {author} {\bibfnamefont {X.}~\bibnamefont {Zhang}}, \bibinfo {author} {\bibfnamefont {Y.}~\bibnamefont {Hu}}, \bibinfo {author} {\bibfnamefont {Z.}~\bibnamefont {Li}}, \bibinfo {author} {\bibfnamefont {Z.}~\bibnamefont {Chen}},\ and\ \bibinfo {author} {\bibfnamefont {S.}~\bibnamefont {Fu}},\ }\bibfield  {title} {\bibinfo {title} {Spin-orbit rabi oscillations in optically synthesized magnetic fields},\ }\href {https://doi.org/10.1038/s41377-023-01238-8} {\bibfield  {journal} {\bibinfo  {journal} {Light: Science {\&} Applications}\ }\textbf {\bibinfo {volume} {12}},\ \bibinfo {pages} {205} (\bibinfo {year} {2023})}\BibitemShut {NoStop}%
\bibitem [{\citenamefont {Boyd}(2008)}]{10.5555/1817101}%
  \BibitemOpen
  \bibfield  {author} {\bibinfo {author} {\bibfnamefont {R.~W.}\ \bibnamefont {Boyd}},\ }\href@noop {} {\emph {\bibinfo {title} {Nonlinear Optics, Third Edition}}},\ \bibinfo {edition} {3rd}\ ed.\ (\bibinfo  {publisher} {Academic Press, Inc.},\ \bibinfo {address} {USA},\ \bibinfo {year} {2008})\BibitemShut {NoStop}%
\bibitem [{\citenamefont {Armstrong}\ \emph {et~al.}(1962)\citenamefont {Armstrong}, \citenamefont {Bloembergen}, \citenamefont {Ducuing},\ and\ \citenamefont {Pershan}}]{PhysRev.127.1918}%
  \BibitemOpen
  \bibfield  {author} {\bibinfo {author} {\bibfnamefont {J.~A.}\ \bibnamefont {Armstrong}}, \bibinfo {author} {\bibfnamefont {N.}~\bibnamefont {Bloembergen}}, \bibinfo {author} {\bibfnamefont {J.}~\bibnamefont {Ducuing}},\ and\ \bibinfo {author} {\bibfnamefont {P.~S.}\ \bibnamefont {Pershan}},\ }\bibfield  {title} {\bibinfo {title} {Interactions between light waves in a nonlinear dielectric},\ }\href {https://doi.org/10.1103/PhysRev.127.1918} {\bibfield  {journal} {\bibinfo  {journal} {Phys. Rev.}\ }\textbf {\bibinfo {volume} {127}},\ \bibinfo {pages} {1918} (\bibinfo {year} {1962})}\BibitemShut {NoStop}%
\bibitem [{\citenamefont {Suchowski}\ \emph {et~al.}(2008)\citenamefont {Suchowski}, \citenamefont {Oron}, \citenamefont {Arie},\ and\ \citenamefont {Silberberg}}]{PhysRevA.78.063821}%
  \BibitemOpen
  \bibfield  {author} {\bibinfo {author} {\bibfnamefont {H.}~\bibnamefont {Suchowski}}, \bibinfo {author} {\bibfnamefont {D.}~\bibnamefont {Oron}}, \bibinfo {author} {\bibfnamefont {A.}~\bibnamefont {Arie}},\ and\ \bibinfo {author} {\bibfnamefont {Y.}~\bibnamefont {Silberberg}},\ }\bibfield  {title} {\bibinfo {title} {Geometrical representation of sum frequency generation and adiabatic frequency conversion},\ }\href {https://doi.org/10.1103/PhysRevA.78.063821} {\bibfield  {journal} {\bibinfo  {journal} {Phys. Rev. A}\ }\textbf {\bibinfo {volume} {78}},\ \bibinfo {pages} {063821} (\bibinfo {year} {2008})}\BibitemShut {NoStop}%
\bibitem [{\citenamefont {Everschor-Sitte}\ and\ \citenamefont {Sitte}(2014)}]{10.1063/1.4870695}%
  \BibitemOpen
  \bibfield  {author} {\bibinfo {author} {\bibfnamefont {K.}~\bibnamefont {Everschor-Sitte}}\ and\ \bibinfo {author} {\bibfnamefont {M.}~\bibnamefont {Sitte}},\ }\bibfield  {title} {\bibinfo {title} {{Real-space Berry phases: Skyrmion soccer (invited)}},\ }\href {https://doi.org/10.1063/1.4870695} {\bibfield  {journal} {\bibinfo  {journal} {Journal of Applied Physics}\ }\textbf {\bibinfo {volume} {115}},\ \bibinfo {pages} {172602} (\bibinfo {year} {2014})}\BibitemShut {NoStop}%
\bibitem [{\citenamefont {Sakurai}\ and\ \citenamefont {Napolitano}(2020)}]{Sakurai_Napolitano_2020}%
  \BibitemOpen
  \bibfield  {author} {\bibinfo {author} {\bibfnamefont {J.~J.}\ \bibnamefont {Sakurai}}\ and\ \bibinfo {author} {\bibfnamefont {J.}~\bibnamefont {Napolitano}},\ }\href@noop {} {\emph {\bibinfo {title} {Modern Quantum Mechanics}}},\ \bibinfo {edition} {3rd}\ ed.\ (\bibinfo  {publisher} {Cambridge University Press},\ \bibinfo {year} {2020})\BibitemShut {NoStop}%
\bibitem [{sup()}]{supplemental}%
  \BibitemOpen
  \href@noop {} {}\bibinfo {note} {See Supplemental Material at [URL] for theoretical derivations, additional simulations, discussion on pseudospin decoherence, and details of the experimental setup, including Refs.~\cite{M_Houe_1995, Chen:16, kibble2004classical}}\BibitemShut {NoStop}%
\bibitem [{\citenamefont {Billy}\ \emph {et~al.}(2008)\citenamefont {Billy}, \citenamefont {Josse}, \citenamefont {Zuo}, \citenamefont {Bernard}, \citenamefont {Hambrecht}, \citenamefont {Lugan}, \citenamefont {Cl{\'e}ment}, \citenamefont {Sanchez-Palencia}, \citenamefont {Bouyer},\ and\ \citenamefont {Aspect}}]{Billy2008}%
  \BibitemOpen
  \bibfield  {author} {\bibinfo {author} {\bibfnamefont {J.}~\bibnamefont {Billy}}, \bibinfo {author} {\bibfnamefont {V.}~\bibnamefont {Josse}}, \bibinfo {author} {\bibfnamefont {Z.}~\bibnamefont {Zuo}}, \bibinfo {author} {\bibfnamefont {A.}~\bibnamefont {Bernard}}, \bibinfo {author} {\bibfnamefont {B.}~\bibnamefont {Hambrecht}}, \bibinfo {author} {\bibfnamefont {P.}~\bibnamefont {Lugan}}, \bibinfo {author} {\bibfnamefont {D.}~\bibnamefont {Cl{\'e}ment}}, \bibinfo {author} {\bibfnamefont {L.}~\bibnamefont {Sanchez-Palencia}}, \bibinfo {author} {\bibfnamefont {P.}~\bibnamefont {Bouyer}},\ and\ \bibinfo {author} {\bibfnamefont {A.}~\bibnamefont {Aspect}},\ }\bibfield  {title} {\bibinfo {title} {Direct observation of anderson localization of matter waves in a controlled disorder},\ }\href {https://doi.org/10.1038/nature07000} {\bibfield  {journal} {\bibinfo  {journal} {Nature}\ }\textbf {\bibinfo {volume} {453}},\ \bibinfo {pages} {891} (\bibinfo {year} {2008})}\BibitemShut {NoStop}%
\bibitem [{\citenamefont {Kondov}\ \emph {et~al.}(2011)\citenamefont {Kondov}, \citenamefont {McGehee}, \citenamefont {Zirbel},\ and\ \citenamefont {DeMarco}}]{doi:10.1126/science.1209019}%
  \BibitemOpen
  \bibfield  {author} {\bibinfo {author} {\bibfnamefont {S.~S.}\ \bibnamefont {Kondov}}, \bibinfo {author} {\bibfnamefont {W.~R.}\ \bibnamefont {McGehee}}, \bibinfo {author} {\bibfnamefont {J.~J.}\ \bibnamefont {Zirbel}},\ and\ \bibinfo {author} {\bibfnamefont {B.}~\bibnamefont {DeMarco}},\ }\bibfield  {title} {\bibinfo {title} {Three-dimensional anderson localization of ultracold matter},\ }\href {https://doi.org/10.1126/science.1209019} {\bibfield  {journal} {\bibinfo  {journal} {Science}\ }\textbf {\bibinfo {volume} {334}},\ \bibinfo {pages} {66} (\bibinfo {year} {2011})}\BibitemShut {NoStop}%
\bibitem [{\citenamefont {Jendrzejewski}\ \emph {et~al.}(2012)\citenamefont {Jendrzejewski}, \citenamefont {Bernard}, \citenamefont {M{\"u}ller}, \citenamefont {Cheinet}, \citenamefont {Josse}, \citenamefont {Piraud}, \citenamefont {Pezz{\'e}}, \citenamefont {Sanchez-Palencia}, \citenamefont {Aspect},\ and\ \citenamefont {Bouyer}}]{Jendrzejewski2012}%
  \BibitemOpen
  \bibfield  {author} {\bibinfo {author} {\bibfnamefont {F.}~\bibnamefont {Jendrzejewski}}, \bibinfo {author} {\bibfnamefont {A.}~\bibnamefont {Bernard}}, \bibinfo {author} {\bibfnamefont {K.}~\bibnamefont {M{\"u}ller}}, \bibinfo {author} {\bibfnamefont {P.}~\bibnamefont {Cheinet}}, \bibinfo {author} {\bibfnamefont {V.}~\bibnamefont {Josse}}, \bibinfo {author} {\bibfnamefont {M.}~\bibnamefont {Piraud}}, \bibinfo {author} {\bibfnamefont {L.}~\bibnamefont {Pezz{\'e}}}, \bibinfo {author} {\bibfnamefont {L.}~\bibnamefont {Sanchez-Palencia}}, \bibinfo {author} {\bibfnamefont {A.}~\bibnamefont {Aspect}},\ and\ \bibinfo {author} {\bibfnamefont {P.}~\bibnamefont {Bouyer}},\ }\bibfield  {title} {\bibinfo {title} {Three-dimensional localization of ultracold atoms in an optical disordered potential},\ }\href {https://doi.org/10.1038/nphys2256} {\bibfield  {journal} {\bibinfo  {journal} {Nature Physics}\ }\textbf {\bibinfo {volume} {8}},\ \bibinfo {pages} {398} (\bibinfo {year} {2012})}\BibitemShut {NoStop}%
\bibitem [{\citenamefont {Liu}\ \emph {et~al.}(2021)\citenamefont {Liu}, \citenamefont {Qing}, \citenamefont {Zhao}, \citenamefont {Zhang}, \citenamefont {Gao}, \citenamefont {Chen},\ and\ \citenamefont {Li}}]{PhysRevLett.127.180601}%
  \BibitemOpen
  \bibfield  {author} {\bibinfo {author} {\bibfnamefont {R.}~\bibnamefont {Liu}}, \bibinfo {author} {\bibfnamefont {B.}~\bibnamefont {Qing}}, \bibinfo {author} {\bibfnamefont {S.}~\bibnamefont {Zhao}}, \bibinfo {author} {\bibfnamefont {P.}~\bibnamefont {Zhang}}, \bibinfo {author} {\bibfnamefont {H.}~\bibnamefont {Gao}}, \bibinfo {author} {\bibfnamefont {S.}~\bibnamefont {Chen}},\ and\ \bibinfo {author} {\bibfnamefont {F.}~\bibnamefont {Li}},\ }\bibfield  {title} {\bibinfo {title} {Generation of non-rayleigh nondiffracting speckles},\ }\href {https://doi.org/10.1103/PhysRevLett.127.180601} {\bibfield  {journal} {\bibinfo  {journal} {Phys. Rev. Lett.}\ }\textbf {\bibinfo {volume} {127}},\ \bibinfo {pages} {180601} (\bibinfo {year} {2021})}\BibitemShut {NoStop}%
\bibitem [{\citenamefont {De~Raedt}\ \emph {et~al.}(1989)\citenamefont {De~Raedt}, \citenamefont {Lagendijk},\ and\ \citenamefont {de~Vries}}]{PhysRevLett.62.47}%
  \BibitemOpen
  \bibfield  {author} {\bibinfo {author} {\bibfnamefont {H.}~\bibnamefont {De~Raedt}}, \bibinfo {author} {\bibfnamefont {A.}~\bibnamefont {Lagendijk}},\ and\ \bibinfo {author} {\bibfnamefont {P.}~\bibnamefont {de~Vries}},\ }\bibfield  {title} {\bibinfo {title} {Transverse localization of light},\ }\href {https://doi.org/10.1103/PhysRevLett.62.47} {\bibfield  {journal} {\bibinfo  {journal} {Phys. Rev. Lett.}\ }\textbf {\bibinfo {volume} {62}},\ \bibinfo {pages} {47} (\bibinfo {year} {1989})}\BibitemShut {NoStop}%
\bibitem [{\citenamefont {Zhang}\ \emph {et~al.}(2021)\citenamefont {Zhang}, \citenamefont {Sheng}, \citenamefont {Zhu}, \citenamefont {Xiao},\ and\ \citenamefont {Krolikowski}}]{Zhang:21}%
  \BibitemOpen
  \bibfield  {author} {\bibinfo {author} {\bibfnamefont {Y.}~\bibnamefont {Zhang}}, \bibinfo {author} {\bibfnamefont {Y.}~\bibnamefont {Sheng}}, \bibinfo {author} {\bibfnamefont {S.}~\bibnamefont {Zhu}}, \bibinfo {author} {\bibfnamefont {M.}~\bibnamefont {Xiao}},\ and\ \bibinfo {author} {\bibfnamefont {W.}~\bibnamefont {Krolikowski}},\ }\bibfield  {title} {\bibinfo {title} {Nonlinear photonic crystals: from 2d to 3d},\ }\href {https://doi.org/10.1364/OPTICA.416619} {\bibfield  {journal} {\bibinfo  {journal} {Optica}\ }\textbf {\bibinfo {volume} {8}},\ \bibinfo {pages} {372} (\bibinfo {year} {2021})}\BibitemShut {NoStop}%
\bibitem [{\citenamefont {Imbrock}\ \emph {et~al.}(2020)\citenamefont {Imbrock}, \citenamefont {Wesemann}, \citenamefont {Kroesen}, \citenamefont {Ayoub},\ and\ \citenamefont {Denz}}]{Imbrock:20}%
  \BibitemOpen
  \bibfield  {author} {\bibinfo {author} {\bibfnamefont {J.}~\bibnamefont {Imbrock}}, \bibinfo {author} {\bibfnamefont {L.}~\bibnamefont {Wesemann}}, \bibinfo {author} {\bibfnamefont {S.}~\bibnamefont {Kroesen}}, \bibinfo {author} {\bibfnamefont {M.}~\bibnamefont {Ayoub}},\ and\ \bibinfo {author} {\bibfnamefont {C.}~\bibnamefont {Denz}},\ }\bibfield  {title} {\bibinfo {title} {Waveguide-integrated three-dimensional quasi-phase-matching structures},\ }\href {https://doi.org/10.1364/OPTICA.7.000028} {\bibfield  {journal} {\bibinfo  {journal} {Optica}\ }\textbf {\bibinfo {volume} {7}},\ \bibinfo {pages} {28} (\bibinfo {year} {2020})}\BibitemShut {NoStop}%
\bibitem [{\citenamefont {Yu}\ \emph {et~al.}(2024)\citenamefont {Yu}, \citenamefont {Liu}, \citenamefont {Xu}, \citenamefont {Zhao},\ and\ \citenamefont {Sheng}}]{Yu:24}%
  \BibitemOpen
  \bibfield  {author} {\bibinfo {author} {\bibfnamefont {C.}~\bibnamefont {Yu}}, \bibinfo {author} {\bibfnamefont {S.}~\bibnamefont {Liu}}, \bibinfo {author} {\bibfnamefont {T.}~\bibnamefont {Xu}}, \bibinfo {author} {\bibfnamefont {R.}~\bibnamefont {Zhao}},\ and\ \bibinfo {author} {\bibfnamefont {Y.}~\bibnamefont {Sheng}},\ }\bibfield  {title} {\bibinfo {title} {Helically twisted nonlinear photonic crystals},\ }\href {https://doi.org/10.1364/OL.532151} {\bibfield  {journal} {\bibinfo  {journal} {Opt. Lett.}\ }\textbf {\bibinfo {volume} {49}},\ \bibinfo {pages} {4745} (\bibinfo {year} {2024})}\BibitemShut {NoStop}%
\bibitem [{\citenamefont {Agrawal}(2013)}]{agrawal2013nonlinear}%
  \BibitemOpen
  \bibfield  {author} {\bibinfo {author} {\bibfnamefont {G.}~\bibnamefont {Agrawal}},\ }\href {https://books.google.co.il/books?id=xNvw-GDVn84C} {\emph {\bibinfo {title} {Nonlinear Fiber Optics}}},\ Optics and Photonics\ (\bibinfo  {publisher} {Elsevier Science},\ \bibinfo {year} {2013})\BibitemShut {NoStop}%
\bibitem [{myr()}]{myref}%
  \BibitemOpen
  \href@noop {} {}\bibinfo {note} {Another disorder measure, based on the modulation phase sampling range, is discussed in SM \cite{supplemental}.}\BibitemShut {Stop}%
\bibitem [{\citenamefont {Hubert}\ and\ \citenamefont {Sch{\"a}fer}(1998)}]{hubert1998magnetic}%
  \BibitemOpen
  \bibfield  {author} {\bibinfo {author} {\bibfnamefont {A.}~\bibnamefont {Hubert}}\ and\ \bibinfo {author} {\bibfnamefont {R.}~\bibnamefont {Sch{\"a}fer}},\ }\href {https://books.google.co.il/books?id=pBE42lLYs-MC} {\emph {\bibinfo {title} {Magnetic Domains: The Analysis of Magnetic Microstructures}}}\ (\bibinfo  {publisher} {Springer},\ \bibinfo {year} {1998})\BibitemShut {NoStop}%
\bibitem [{\citenamefont {Sun}\ \emph {et~al.}(2021)\citenamefont {Sun}, \citenamefont {Song}, \citenamefont {Anderson}, \citenamefont {Brunner}, \citenamefont {F{\"o}rster}, \citenamefont {Shalomayeva}, \citenamefont {Taniguchi}, \citenamefont {Watanabe}, \citenamefont {Gr{\"a}fe}, \citenamefont {St{\"o}hr}, \citenamefont {Xu},\ and\ \citenamefont {Wrachtrup}}]{Sun2021}%
  \BibitemOpen
  \bibfield  {author} {\bibinfo {author} {\bibfnamefont {Q.-C.}\ \bibnamefont {Sun}}, \bibinfo {author} {\bibfnamefont {T.}~\bibnamefont {Song}}, \bibinfo {author} {\bibfnamefont {E.}~\bibnamefont {Anderson}}, \bibinfo {author} {\bibfnamefont {A.}~\bibnamefont {Brunner}}, \bibinfo {author} {\bibfnamefont {J.}~\bibnamefont {F{\"o}rster}}, \bibinfo {author} {\bibfnamefont {T.}~\bibnamefont {Shalomayeva}}, \bibinfo {author} {\bibfnamefont {T.}~\bibnamefont {Taniguchi}}, \bibinfo {author} {\bibfnamefont {K.}~\bibnamefont {Watanabe}}, \bibinfo {author} {\bibfnamefont {J.}~\bibnamefont {Gr{\"a}fe}}, \bibinfo {author} {\bibfnamefont {R.}~\bibnamefont {St{\"o}hr}}, \bibinfo {author} {\bibfnamefont {X.}~\bibnamefont {Xu}},\ and\ \bibinfo {author} {\bibfnamefont {J.}~\bibnamefont {Wrachtrup}},\ }\bibfield  {title} {\bibinfo {title} {Magnetic domains and domain wall pinning in atomically thin crbr3 revealed by nanoscale imaging},\ }\href {https://doi.org/10.1038/s41467-021-22239-4} {\bibfield  {journal} {\bibinfo
  {journal} {Nature Communications}\ }\textbf {\bibinfo {volume} {12}},\ \bibinfo {pages} {1989} (\bibinfo {year} {2021})}\BibitemShut {NoStop}%
\bibitem [{\citenamefont {{De Raedt}}(1987)}]{DERAEDT19871}%
  \BibitemOpen
  \bibfield  {author} {\bibinfo {author} {\bibfnamefont {H.}~\bibnamefont {{De Raedt}}},\ }\bibfield  {title} {\bibinfo {title} {Product formula algorithms for solving the time dependent schrödinger equation},\ }\href {https://doi.org/https://doi.org/10.1016/0167-7977(87)90002-5} {\bibfield  {journal} {\bibinfo  {journal} {Computer Physics Reports}\ }\textbf {\bibinfo {volume} {7}},\ \bibinfo {pages} {1} (\bibinfo {year} {1987})}\BibitemShut {NoStop}%
\bibitem [{\citenamefont {Murphy}\ \emph {et~al.}(2011)\citenamefont {Murphy}, \citenamefont {Wortis},\ and\ \citenamefont {Atkinson}}]{PhysRevB.83.184206}%
  \BibitemOpen
  \bibfield  {author} {\bibinfo {author} {\bibfnamefont {N.~C.}\ \bibnamefont {Murphy}}, \bibinfo {author} {\bibfnamefont {R.}~\bibnamefont {Wortis}},\ and\ \bibinfo {author} {\bibfnamefont {W.~A.}\ \bibnamefont {Atkinson}},\ }\bibfield  {title} {\bibinfo {title} {Generalized inverse participation ratio as a possible measure of localization for interacting systems},\ }\href {https://doi.org/10.1103/PhysRevB.83.184206} {\bibfield  {journal} {\bibinfo  {journal} {Phys. Rev. B}\ }\textbf {\bibinfo {volume} {83}},\ \bibinfo {pages} {184206} (\bibinfo {year} {2011})}\BibitemShut {NoStop}%
\bibitem [{\citenamefont {Wang}\ \emph {et~al.}(2024)\citenamefont {Wang}, \citenamefont {Fu}, \citenamefont {Konotop}, \citenamefont {Kartashov},\ and\ \citenamefont {Ye}}]{Wang2024}%
  \BibitemOpen
  \bibfield  {author} {\bibinfo {author} {\bibfnamefont {P.}~\bibnamefont {Wang}}, \bibinfo {author} {\bibfnamefont {Q.}~\bibnamefont {Fu}}, \bibinfo {author} {\bibfnamefont {V.~V.}\ \bibnamefont {Konotop}}, \bibinfo {author} {\bibfnamefont {Y.~V.}\ \bibnamefont {Kartashov}},\ and\ \bibinfo {author} {\bibfnamefont {F.}~\bibnamefont {Ye}},\ }\bibfield  {title} {\bibinfo {title} {Observation of localization of light in linear photonic quasicrystals with diverse rotational symmetries},\ }\href {https://doi.org/10.1038/s41566-023-01350-6} {\bibfield  {journal} {\bibinfo  {journal} {Nature Photonics}\ }\textbf {\bibinfo {volume} {18}},\ \bibinfo {pages} {224} (\bibinfo {year} {2024})}\BibitemShut {NoStop}%
\bibitem [{()}]{myref1}%
  \BibitemOpen
  \href@noop {} {}\bibinfo {note} {The localization length from the exponential fit is $\xi_{\text{loc}} \approx 104 \ \mu\text{m}$, yielding an FWHM of $72 \ \mu\text{m}$, over twice as narrow as the $\sim 181 \ \mu\text{m}$ FWHM observed in the absence of disorder.}\BibitemShut {Stop}%
\bibitem [{\citenamefont {Yoo}\ \emph {et~al.}(1989)\citenamefont {Yoo}, \citenamefont {Lee}, \citenamefont {Takiguchi},\ and\ \citenamefont {Alfano}}]{Yoo:89}%
  \BibitemOpen
  \bibfield  {author} {\bibinfo {author} {\bibfnamefont {K.~M.}\ \bibnamefont {Yoo}}, \bibinfo {author} {\bibfnamefont {S.}~\bibnamefont {Lee}}, \bibinfo {author} {\bibfnamefont {Y.}~\bibnamefont {Takiguchi}},\ and\ \bibinfo {author} {\bibfnamefont {R.~R.}\ \bibnamefont {Alfano}},\ }\bibfield  {title} {\bibinfo {title} {Search for the effect of weak photon localization in second-harmonic waves generated in a disordered anisotropic nonlinear medium},\ }\href {https://doi.org/10.1364/OL.14.000800} {\bibfield  {journal} {\bibinfo  {journal} {Opt. Lett.}\ }\textbf {\bibinfo {volume} {14}},\ \bibinfo {pages} {800} (\bibinfo {year} {1989})}\BibitemShut {NoStop}%
\bibitem [{foo()}]{footnote4}%
  \BibitemOpen
  \href@noop {} {}\bibinfo {note} {In terms of FWHM, under disorder, the fitted profile widths in the simulation (Fig.~\ref{fig:figure4}(b)) and experiment (Fig.~\ref{fig:figure4}(e)) are $84 \ \mu$m and $89 \ \mu$m, respectively, nearly half the widths observed without disorder ($162 \ \mu$m and $158 \ \mu$m in Figs.~\ref{fig:figure4}(a) and ~\ref{fig:figure4}(d), respectively). This demonstrates significant beam narrowing due to localization.}\BibitemShut {Stop}%
\bibitem [{\citenamefont {Villain}(1979)}]{Villain1979}%
  \BibitemOpen
  \bibfield  {author} {\bibinfo {author} {\bibfnamefont {J.}~\bibnamefont {Villain}},\ }\bibfield  {title} {\bibinfo {title} {Insulating spin glasses},\ }\href {https://doi.org/10.1007/BF01325811} {\bibfield  {journal} {\bibinfo  {journal} {Zeitschrift f{\"u}r Physik B Condensed Matter}\ }\textbf {\bibinfo {volume} {33}},\ \bibinfo {pages} {31} (\bibinfo {year} {1979})}\BibitemShut {NoStop}%
\bibitem [{\citenamefont {Tarnopolsky}\ \emph {et~al.}(2020)\citenamefont {Tarnopolsky}, \citenamefont {Li}, \citenamefont {Joshi},\ and\ \citenamefont {Sachdev}}]{PhysRevB.101.205106}%
  \BibitemOpen
  \bibfield  {author} {\bibinfo {author} {\bibfnamefont {G.}~\bibnamefont {Tarnopolsky}}, \bibinfo {author} {\bibfnamefont {C.}~\bibnamefont {Li}}, \bibinfo {author} {\bibfnamefont {D.~G.}\ \bibnamefont {Joshi}},\ and\ \bibinfo {author} {\bibfnamefont {S.}~\bibnamefont {Sachdev}},\ }\bibfield  {title} {\bibinfo {title} {Metal-insulator transition in a random hubbard model},\ }\href {https://doi.org/10.1103/PhysRevB.101.205106} {\bibfield  {journal} {\bibinfo  {journal} {Phys. Rev. B}\ }\textbf {\bibinfo {volume} {101}},\ \bibinfo {pages} {205106} (\bibinfo {year} {2020})}\BibitemShut {NoStop}%
\bibitem [{\citenamefont {Nishimori}(2022)}]{Nishimori_2022}%
  \BibitemOpen
  \bibfield  {author} {\bibinfo {author} {\bibfnamefont {H.}~\bibnamefont {Nishimori}},\ }\bibfield  {title} {\bibinfo {title} {Analyticity of the energy in an ising spin glass with correlated disorder},\ }\href {https://doi.org/10.1088/1751-8121/ac44ef} {\bibfield  {journal} {\bibinfo  {journal} {Journal of Physics A: Mathematical and Theoretical}\ }\textbf {\bibinfo {volume} {55}},\ \bibinfo {pages} {045001} (\bibinfo {year} {2022})}\BibitemShut {NoStop}%
\bibitem [{\citenamefont {Krempask{\'y}}\ \emph {et~al.}(2023)\citenamefont {Krempask{\'y}}, \citenamefont {Springholz}, \citenamefont {D'Souza}, \citenamefont {Caha}, \citenamefont {Gmitra}, \citenamefont {Ney}, \citenamefont {Vaz}, \citenamefont {Piamonteze}, \citenamefont {Fanciulli}, \citenamefont {Kriegner}, \citenamefont {Krieger}, \citenamefont {Prokscha}, \citenamefont {Salman}, \citenamefont {Min{\'a}r},\ and\ \citenamefont {Dil}}]{Krempaský2023}%
  \BibitemOpen
  \bibfield  {author} {\bibinfo {author} {\bibfnamefont {J.}~\bibnamefont {Krempask{\'y}}}, \bibinfo {author} {\bibfnamefont {G.}~\bibnamefont {Springholz}}, \bibinfo {author} {\bibfnamefont {S.~W.}\ \bibnamefont {D'Souza}}, \bibinfo {author} {\bibfnamefont {O.}~\bibnamefont {Caha}}, \bibinfo {author} {\bibfnamefont {M.}~\bibnamefont {Gmitra}}, \bibinfo {author} {\bibfnamefont {A.}~\bibnamefont {Ney}}, \bibinfo {author} {\bibfnamefont {C.~A.~F.}\ \bibnamefont {Vaz}}, \bibinfo {author} {\bibfnamefont {C.}~\bibnamefont {Piamonteze}}, \bibinfo {author} {\bibfnamefont {M.}~\bibnamefont {Fanciulli}}, \bibinfo {author} {\bibfnamefont {D.}~\bibnamefont {Kriegner}}, \bibinfo {author} {\bibfnamefont {J.~A.}\ \bibnamefont {Krieger}}, \bibinfo {author} {\bibfnamefont {T.}~\bibnamefont {Prokscha}}, \bibinfo {author} {\bibfnamefont {Z.}~\bibnamefont {Salman}}, \bibinfo {author} {\bibfnamefont {J.}~\bibnamefont {Min{\'a}r}},\ and\ \bibinfo {author} {\bibfnamefont {J.~H.}\ \bibnamefont {Dil}},\ }\bibfield  {title} {\bibinfo
  {title} {Efficient magnetic switching in a correlated spin glass},\ }\href {https://doi.org/10.1038/s41467-023-41718-4} {\bibfield  {journal} {\bibinfo  {journal} {Nature Communications}\ }\textbf {\bibinfo {volume} {14}},\ \bibinfo {pages} {6127} (\bibinfo {year} {2023})}\BibitemShut {NoStop}%
\bibitem [{\citenamefont {M{\'e}zard}(2024)}]{Mézard2024}%
  \BibitemOpen
  \bibfield  {author} {\bibinfo {author} {\bibfnamefont {M.}~\bibnamefont {M{\'e}zard}},\ }\bibfield  {title} {\bibinfo {title} {Spin glass theory and its new challenge: structured disorder},\ }\href {https://doi.org/10.1007/s12648-023-03029-8} {\bibfield  {journal} {\bibinfo  {journal} {Indian Journal of Physics}\ }\textbf {\bibinfo {volume} {98}},\ \bibinfo {pages} {3757} (\bibinfo {year} {2024})}\BibitemShut {NoStop}%
\bibitem [{\citenamefont {Skj{\ae}rv{\o}}\ \emph {et~al.}(2020)\citenamefont {Skj{\ae}rv{\o}}, \citenamefont {Marrows}, \citenamefont {Stamps},\ and\ \citenamefont {Heyderman}}]{Skjærvø2020}%
  \BibitemOpen
  \bibfield  {author} {\bibinfo {author} {\bibfnamefont {S.~H.}\ \bibnamefont {Skj{\ae}rv{\o}}}, \bibinfo {author} {\bibfnamefont {C.~H.}\ \bibnamefont {Marrows}}, \bibinfo {author} {\bibfnamefont {R.~L.}\ \bibnamefont {Stamps}},\ and\ \bibinfo {author} {\bibfnamefont {L.~J.}\ \bibnamefont {Heyderman}},\ }\bibfield  {title} {\bibinfo {title} {Advances in artificial spin ice},\ }\href {https://doi.org/10.1038/s42254-019-0118-3} {\bibfield  {journal} {\bibinfo  {journal} {Nature Reviews Physics}\ }\textbf {\bibinfo {volume} {2}},\ \bibinfo {pages} {13} (\bibinfo {year} {2020})}\BibitemShut {NoStop}%
\bibitem [{\citenamefont {Lendinez}\ and\ \citenamefont {Jungfleisch}(2019)}]{Lendinez_2020}%
  \BibitemOpen
  \bibfield  {author} {\bibinfo {author} {\bibfnamefont {S.}~\bibnamefont {Lendinez}}\ and\ \bibinfo {author} {\bibfnamefont {M.~B.}\ \bibnamefont {Jungfleisch}},\ }\bibfield  {title} {\bibinfo {title} {Magnetization dynamics in artificial spin ice},\ }\href {https://doi.org/10.1088/1361-648X/ab3e78} {\bibfield  {journal} {\bibinfo  {journal} {Journal of Physics: Condensed Matter}\ }\textbf {\bibinfo {volume} {32}},\ \bibinfo {pages} {013001} (\bibinfo {year} {2019})}\BibitemShut {NoStop}%
\bibitem [{\citenamefont {Tserkovnyak}\ and\ \citenamefont {Ochoa}(2017)}]{PhysRevB.96.100402}%
  \BibitemOpen
  \bibfield  {author} {\bibinfo {author} {\bibfnamefont {Y.}~\bibnamefont {Tserkovnyak}}\ and\ \bibinfo {author} {\bibfnamefont {H.}~\bibnamefont {Ochoa}},\ }\bibfield  {title} {\bibinfo {title} {Generalized boundary conditions for spin transfer},\ }\href {https://doi.org/10.1103/PhysRevB.96.100402} {\bibfield  {journal} {\bibinfo  {journal} {Phys. Rev. B}\ }\textbf {\bibinfo {volume} {96}},\ \bibinfo {pages} {100402} (\bibinfo {year} {2017})}\BibitemShut {NoStop}%
\bibitem [{\citenamefont {Mondal}\ and\ \citenamefont {Das}(2022)}]{Mondal:22}%
  \BibitemOpen
  \bibfield  {author} {\bibinfo {author} {\bibfnamefont {A.}~\bibnamefont {Mondal}}\ and\ \bibinfo {author} {\bibfnamefont {R.}~\bibnamefont {Das}},\ }\bibfield  {title} {\bibinfo {title} {Experimental evidence of a pump-wavefront-induced stern--gerlach-like splitting in optical parametric generators},\ }\href {https://doi.org/10.1364/OL.460995} {\bibfield  {journal} {\bibinfo  {journal} {Opt. Lett.}\ }\textbf {\bibinfo {volume} {47}},\ \bibinfo {pages} {3668} (\bibinfo {year} {2022})}\BibitemShut {NoStop}%
\bibitem [{\citenamefont {Ding}\ \emph {et~al.}(2020)\citenamefont {Ding}, \citenamefont {Heberle}, \citenamefont {Harrington}, \citenamefont {Flemens}, \citenamefont {Chang}, \citenamefont {Birks},\ and\ \citenamefont {Moses}}]{PhysRevLett.124.153902}%
  \BibitemOpen
  \bibfield  {author} {\bibinfo {author} {\bibfnamefont {X.}~\bibnamefont {Ding}}, \bibinfo {author} {\bibfnamefont {D.}~\bibnamefont {Heberle}}, \bibinfo {author} {\bibfnamefont {K.}~\bibnamefont {Harrington}}, \bibinfo {author} {\bibfnamefont {N.}~\bibnamefont {Flemens}}, \bibinfo {author} {\bibfnamefont {W.-Z.}\ \bibnamefont {Chang}}, \bibinfo {author} {\bibfnamefont {T.~A.}\ \bibnamefont {Birks}},\ and\ \bibinfo {author} {\bibfnamefont {J.}~\bibnamefont {Moses}},\ }\bibfield  {title} {\bibinfo {title} {Observation of rapid adiabatic passage in optical four-wave mixing},\ }\href {https://doi.org/10.1103/PhysRevLett.124.153902} {\bibfield  {journal} {\bibinfo  {journal} {Phys. Rev. Lett.}\ }\textbf {\bibinfo {volume} {124}},\ \bibinfo {pages} {153902} (\bibinfo {year} {2020})}\BibitemShut {NoStop}%
\bibitem [{\citenamefont {Li}\ \emph {et~al.}(2021)\citenamefont {Li}, \citenamefont {L\"{u}}, \citenamefont {Fu},\ and\ \citenamefont {Arie}}]{Li:21}%
  \BibitemOpen
  \bibfield  {author} {\bibinfo {author} {\bibfnamefont {Y.}~\bibnamefont {Li}}, \bibinfo {author} {\bibfnamefont {J.}~\bibnamefont {L\"{u}}}, \bibinfo {author} {\bibfnamefont {S.}~\bibnamefont {Fu}},\ and\ \bibinfo {author} {\bibfnamefont {A.}~\bibnamefont {Arie}},\ }\bibfield  {title} {\bibinfo {title} {Geometric representation and the adiabatic geometric phase in four-wave mixing processes},\ }\href {https://doi.org/10.1364/OE.416186} {\bibfield  {journal} {\bibinfo  {journal} {Opt. Express}\ }\textbf {\bibinfo {volume} {29}},\ \bibinfo {pages} {7288} (\bibinfo {year} {2021})}\BibitemShut {NoStop}%
\bibitem [{\citenamefont {Bashan}\ \emph {et~al.}(2024)\citenamefont {Bashan}, \citenamefont {Eyal}, \citenamefont {Tur},\ and\ \citenamefont {Arie}}]{Bashan:24}%
  \BibitemOpen
  \bibfield  {author} {\bibinfo {author} {\bibfnamefont {G.}~\bibnamefont {Bashan}}, \bibinfo {author} {\bibfnamefont {A.}~\bibnamefont {Eyal}}, \bibinfo {author} {\bibfnamefont {M.}~\bibnamefont {Tur}},\ and\ \bibinfo {author} {\bibfnamefont {A.}~\bibnamefont {Arie}},\ }\bibfield  {title} {\bibinfo {title} {All-optical stern-gerlach effect in the time domain},\ }\href {https://doi.org/10.1364/OE.510722} {\bibfield  {journal} {\bibinfo  {journal} {Opt. Express}\ }\textbf {\bibinfo {volume} {32}},\ \bibinfo {pages} {9589} (\bibinfo {year} {2024})}\BibitemShut {NoStop}%
\bibitem [{\citenamefont {Zhu}\ \emph {et~al.}(2024)\citenamefont {Zhu}, \citenamefont {Liang}, \citenamefont {Xie}, \citenamefont {Zheng}, \citenamefont {Zhong}, \citenamefont {Tang}, \citenamefont {Yu},\ and\ \citenamefont {Chen}}]{https://doi.org/10.1002/lpor.202301030}%
  \BibitemOpen
  \bibfield  {author} {\bibinfo {author} {\bibfnamefont {W.}~\bibnamefont {Zhu}}, \bibinfo {author} {\bibfnamefont {X.}~\bibnamefont {Liang}}, \bibinfo {author} {\bibfnamefont {W.}~\bibnamefont {Xie}}, \bibinfo {author} {\bibfnamefont {H.}~\bibnamefont {Zheng}}, \bibinfo {author} {\bibfnamefont {Y.}~\bibnamefont {Zhong}}, \bibinfo {author} {\bibfnamefont {J.}~\bibnamefont {Tang}}, \bibinfo {author} {\bibfnamefont {J.}~\bibnamefont {Yu}},\ and\ \bibinfo {author} {\bibfnamefont {Z.}~\bibnamefont {Chen}},\ }\bibfield  {title} {\bibinfo {title} {Optical stern–gerlach effect in periodically poled electro-optical crystals},\ }\href {https://doi.org/https://doi.org/10.1002/lpor.202301030} {\bibfield  {journal} {\bibinfo  {journal} {Laser \& Photonics Reviews}\ }\textbf {\bibinfo {volume} {18}},\ \bibinfo {pages} {2301030} (\bibinfo {year} {2024})}\BibitemShut {NoStop}%
\bibitem [{\citenamefont {Liu}\ \emph {et~al.}(2024)\citenamefont {Liu}, \citenamefont {Zeng}, \citenamefont {Lin}, \citenamefont {Hu}, \citenamefont {Li}, \citenamefont {Chen},\ and\ \citenamefont {Fu}}]{Liu:24}%
  \BibitemOpen
  \bibfield  {author} {\bibinfo {author} {\bibfnamefont {G.}~\bibnamefont {Liu}}, \bibinfo {author} {\bibfnamefont {Z.}~\bibnamefont {Zeng}}, \bibinfo {author} {\bibfnamefont {H.}~\bibnamefont {Lin}}, \bibinfo {author} {\bibfnamefont {Y.}~\bibnamefont {Hu}}, \bibinfo {author} {\bibfnamefont {Z.}~\bibnamefont {Li}}, \bibinfo {author} {\bibfnamefont {Z.}~\bibnamefont {Chen}},\ and\ \bibinfo {author} {\bibfnamefont {S.}~\bibnamefont {Fu}},\ }\bibfield  {title} {\bibinfo {title} {Electrically engineering synthetic magnetic fields for polarized photons},\ }\href {https://doi.org/10.1364/OPTICA.527811} {\bibfield  {journal} {\bibinfo  {journal} {Optica}\ }\textbf {\bibinfo {volume} {11}},\ \bibinfo {pages} {980} (\bibinfo {year} {2024})}\BibitemShut {NoStop}%
\bibitem [{\citenamefont {Li}\ \emph {et~al.}(2024)\citenamefont {Li}, \citenamefont {Wang}, \citenamefont {Hu},\ and\ \citenamefont {Xu}}]{https://doi.org/10.1002/lpor.202301055}%
  \BibitemOpen
  \bibfield  {author} {\bibinfo {author} {\bibfnamefont {J.}~\bibnamefont {Li}}, \bibinfo {author} {\bibfnamefont {Z.}~\bibnamefont {Wang}}, \bibinfo {author} {\bibfnamefont {Y.}~\bibnamefont {Hu}},\ and\ \bibinfo {author} {\bibfnamefont {J.}~\bibnamefont {Xu}},\ }\bibfield  {title} {\bibinfo {title} {Splitting of optical spatial modes via stern-gerlach effect},\ }\href {https://doi.org/https://doi.org/10.1002/lpor.202301055} {\bibfield  {journal} {\bibinfo  {journal} {Laser \& Photonics Reviews}\ }\textbf {\bibinfo {volume} {18}},\ \bibinfo {pages} {2301055} (\bibinfo {year} {2024})}\BibitemShut {NoStop}%
\bibitem [{\citenamefont {Kobayashi}\ \emph {et~al.}(2016)\citenamefont {Kobayashi}, \citenamefont {Ikuta}, \citenamefont {Yasui}, \citenamefont {Miki}, \citenamefont {Yamashita}, \citenamefont {Terai}, \citenamefont {Yamamoto}, \citenamefont {Koashi},\ and\ \citenamefont {Imoto}}]{Kobayashi2016}%
  \BibitemOpen
  \bibfield  {author} {\bibinfo {author} {\bibfnamefont {T.}~\bibnamefont {Kobayashi}}, \bibinfo {author} {\bibfnamefont {R.}~\bibnamefont {Ikuta}}, \bibinfo {author} {\bibfnamefont {S.}~\bibnamefont {Yasui}}, \bibinfo {author} {\bibfnamefont {S.}~\bibnamefont {Miki}}, \bibinfo {author} {\bibfnamefont {T.}~\bibnamefont {Yamashita}}, \bibinfo {author} {\bibfnamefont {H.}~\bibnamefont {Terai}}, \bibinfo {author} {\bibfnamefont {T.}~\bibnamefont {Yamamoto}}, \bibinfo {author} {\bibfnamefont {M.}~\bibnamefont {Koashi}},\ and\ \bibinfo {author} {\bibfnamefont {N.}~\bibnamefont {Imoto}},\ }\bibfield  {title} {\bibinfo {title} {Frequency-domain hong--ou--mandel interference},\ }\href {https://doi.org/10.1038/nphoton.2016.74} {\bibfield  {journal} {\bibinfo  {journal} {Nature Photonics}\ }\textbf {\bibinfo {volume} {10}},\ \bibinfo {pages} {441} (\bibinfo {year} {2016})}\BibitemShut {NoStop}%
\bibitem [{\citenamefont {Kues}\ \emph {et~al.}(2019)\citenamefont {Kues}, \citenamefont {Reimer}, \citenamefont {Lukens}, \citenamefont {Munro}, \citenamefont {Weiner}, \citenamefont {Moss},\ and\ \citenamefont {Morandotti}}]{Kues2019}%
  \BibitemOpen
  \bibfield  {author} {\bibinfo {author} {\bibfnamefont {M.}~\bibnamefont {Kues}}, \bibinfo {author} {\bibfnamefont {C.}~\bibnamefont {Reimer}}, \bibinfo {author} {\bibfnamefont {J.~M.}\ \bibnamefont {Lukens}}, \bibinfo {author} {\bibfnamefont {W.~J.}\ \bibnamefont {Munro}}, \bibinfo {author} {\bibfnamefont {A.~M.}\ \bibnamefont {Weiner}}, \bibinfo {author} {\bibfnamefont {D.~J.}\ \bibnamefont {Moss}},\ and\ \bibinfo {author} {\bibfnamefont {R.}~\bibnamefont {Morandotti}},\ }\bibfield  {title} {\bibinfo {title} {Quantum optical microcombs},\ }\href {https://doi.org/10.1038/s41566-019-0363-0} {\bibfield  {journal} {\bibinfo  {journal} {Nature Photonics}\ }\textbf {\bibinfo {volume} {13}},\ \bibinfo {pages} {170} (\bibinfo {year} {2019})}\BibitemShut {NoStop}%
\bibitem [{\citenamefont {Joshi}\ \emph {et~al.}(2020)\citenamefont {Joshi}, \citenamefont {Farsi}, \citenamefont {Dutt}, \citenamefont {Kim}, \citenamefont {Ji}, \citenamefont {Zhao}, \citenamefont {Bishop}, \citenamefont {Lipson},\ and\ \citenamefont {Gaeta}}]{PhysRevLett.124.143601}%
  \BibitemOpen
  \bibfield  {author} {\bibinfo {author} {\bibfnamefont {C.}~\bibnamefont {Joshi}}, \bibinfo {author} {\bibfnamefont {A.}~\bibnamefont {Farsi}}, \bibinfo {author} {\bibfnamefont {A.}~\bibnamefont {Dutt}}, \bibinfo {author} {\bibfnamefont {B.~Y.}\ \bibnamefont {Kim}}, \bibinfo {author} {\bibfnamefont {X.}~\bibnamefont {Ji}}, \bibinfo {author} {\bibfnamefont {Y.}~\bibnamefont {Zhao}}, \bibinfo {author} {\bibfnamefont {A.~M.}\ \bibnamefont {Bishop}}, \bibinfo {author} {\bibfnamefont {M.}~\bibnamefont {Lipson}},\ and\ \bibinfo {author} {\bibfnamefont {A.~L.}\ \bibnamefont {Gaeta}},\ }\bibfield  {title} {\bibinfo {title} {Frequency-domain quantum interference with correlated photons from an integrated microresonator},\ }\href {https://doi.org/10.1103/PhysRevLett.124.143601} {\bibfield  {journal} {\bibinfo  {journal} {Phys. Rev. Lett.}\ }\textbf {\bibinfo {volume} {124}},\ \bibinfo {pages} {143601} (\bibinfo {year} {2020})}\BibitemShut {NoStop}%
\bibitem [{\citenamefont {Lu}\ \emph {et~al.}(2023)\citenamefont {Lu}, \citenamefont {Liscidini}, \citenamefont {Gaeta}, \citenamefont {Weiner},\ and\ \citenamefont {Lukens}}]{Lu:23}%
  \BibitemOpen
  \bibfield  {author} {\bibinfo {author} {\bibfnamefont {H.-H.}\ \bibnamefont {Lu}}, \bibinfo {author} {\bibfnamefont {M.}~\bibnamefont {Liscidini}}, \bibinfo {author} {\bibfnamefont {A.~L.}\ \bibnamefont {Gaeta}}, \bibinfo {author} {\bibfnamefont {A.~M.}\ \bibnamefont {Weiner}},\ and\ \bibinfo {author} {\bibfnamefont {J.~M.}\ \bibnamefont {Lukens}},\ }\bibfield  {title} {\bibinfo {title} {Frequency-bin photonic quantum information},\ }\href {https://doi.org/10.1364/OPTICA.506096} {\bibfield  {journal} {\bibinfo  {journal} {Optica}\ }\textbf {\bibinfo {volume} {10}},\ \bibinfo {pages} {1655} (\bibinfo {year} {2023})}\BibitemShut {NoStop}%
\bibitem [{\citenamefont {Aharonov}\ \emph {et~al.}(1993)\citenamefont {Aharonov}, \citenamefont {Davidovich},\ and\ \citenamefont {Zagury}}]{PhysRevA.48.1687}%
  \BibitemOpen
  \bibfield  {author} {\bibinfo {author} {\bibfnamefont {Y.}~\bibnamefont {Aharonov}}, \bibinfo {author} {\bibfnamefont {L.}~\bibnamefont {Davidovich}},\ and\ \bibinfo {author} {\bibfnamefont {N.}~\bibnamefont {Zagury}},\ }\bibfield  {title} {\bibinfo {title} {Quantum random walks},\ }\href {https://doi.org/10.1103/PhysRevA.48.1687} {\bibfield  {journal} {\bibinfo  {journal} {Phys. Rev. A}\ }\textbf {\bibinfo {volume} {48}},\ \bibinfo {pages} {1687} (\bibinfo {year} {1993})}\BibitemShut {NoStop}%
\bibitem [{\citenamefont {Lahini}\ \emph {et~al.}(2010)\citenamefont {Lahini}, \citenamefont {Bromberg}, \citenamefont {Christodoulides},\ and\ \citenamefont {Silberberg}}]{PhysRevLett.105.163905}%
  \BibitemOpen
  \bibfield  {author} {\bibinfo {author} {\bibfnamefont {Y.}~\bibnamefont {Lahini}}, \bibinfo {author} {\bibfnamefont {Y.}~\bibnamefont {Bromberg}}, \bibinfo {author} {\bibfnamefont {D.~N.}\ \bibnamefont {Christodoulides}},\ and\ \bibinfo {author} {\bibfnamefont {Y.}~\bibnamefont {Silberberg}},\ }\bibfield  {title} {\bibinfo {title} {Quantum correlations in two-particle anderson localization},\ }\href {https://doi.org/10.1103/PhysRevLett.105.163905} {\bibfield  {journal} {\bibinfo  {journal} {Phys. Rev. Lett.}\ }\textbf {\bibinfo {volume} {105}},\ \bibinfo {pages} {163905} (\bibinfo {year} {2010})}\BibitemShut {NoStop}%
\bibitem [{\citenamefont {Bromberg}\ \emph {et~al.}(2009)\citenamefont {Bromberg}, \citenamefont {Lahini}, \citenamefont {Morandotti},\ and\ \citenamefont {Silberberg}}]{PhysRevLett.102.253904}%
  \BibitemOpen
  \bibfield  {author} {\bibinfo {author} {\bibfnamefont {Y.}~\bibnamefont {Bromberg}}, \bibinfo {author} {\bibfnamefont {Y.}~\bibnamefont {Lahini}}, \bibinfo {author} {\bibfnamefont {R.}~\bibnamefont {Morandotti}},\ and\ \bibinfo {author} {\bibfnamefont {Y.}~\bibnamefont {Silberberg}},\ }\bibfield  {title} {\bibinfo {title} {Quantum and classical correlations in waveguide lattices},\ }\href {https://doi.org/10.1103/PhysRevLett.102.253904} {\bibfield  {journal} {\bibinfo  {journal} {Phys. Rev. Lett.}\ }\textbf {\bibinfo {volume} {102}},\ \bibinfo {pages} {253904} (\bibinfo {year} {2009})}\BibitemShut {NoStop}%
\bibitem [{\citenamefont {Peruzzo}\ \emph {et~al.}(2010)\citenamefont {Peruzzo}, \citenamefont {Lobino}, \citenamefont {Matthews}, \citenamefont {Matsuda}, \citenamefont {Politi}, \citenamefont {Poulios}, \citenamefont {Zhou}, \citenamefont {Lahini}, \citenamefont {Ismail}, \citenamefont {Wörhoff}, \citenamefont {Bromberg}, \citenamefont {Silberberg}, \citenamefont {Thompson},\ and\ \citenamefont {OBrien}}]{doi:10.1126/science.1193515}%
  \BibitemOpen
  \bibfield  {author} {\bibinfo {author} {\bibfnamefont {A.}~\bibnamefont {Peruzzo}}, \bibinfo {author} {\bibfnamefont {M.}~\bibnamefont {Lobino}}, \bibinfo {author} {\bibfnamefont {J.~C.~F.}\ \bibnamefont {Matthews}}, \bibinfo {author} {\bibfnamefont {N.}~\bibnamefont {Matsuda}}, \bibinfo {author} {\bibfnamefont {A.}~\bibnamefont {Politi}}, \bibinfo {author} {\bibfnamefont {K.}~\bibnamefont {Poulios}}, \bibinfo {author} {\bibfnamefont {X.-Q.}\ \bibnamefont {Zhou}}, \bibinfo {author} {\bibfnamefont {Y.}~\bibnamefont {Lahini}}, \bibinfo {author} {\bibfnamefont {N.}~\bibnamefont {Ismail}}, \bibinfo {author} {\bibfnamefont {K.}~\bibnamefont {Wörhoff}}, \bibinfo {author} {\bibfnamefont {Y.}~\bibnamefont {Bromberg}}, \bibinfo {author} {\bibfnamefont {Y.}~\bibnamefont {Silberberg}}, \bibinfo {author} {\bibfnamefont {M.~G.}\ \bibnamefont {Thompson}},\ and\ \bibinfo {author} {\bibfnamefont {J.~L.}\ \bibnamefont {OBrien}},\ }\bibfield  {title} {\bibinfo {title} {Quantum walks of correlated photons},\ }\href
  {https://doi.org/10.1126/science.1193515} {\bibfield  {journal} {\bibinfo  {journal} {Science}\ }\textbf {\bibinfo {volume} {329}},\ \bibinfo {pages} {1500} (\bibinfo {year} {2010})}\BibitemShut {NoStop}%
\bibitem [{\citenamefont {Houe}\ and\ \citenamefont {Townsend}(1995)}]{M_Houe_1995}%
  \BibitemOpen
  \bibfield  {author} {\bibinfo {author} {\bibfnamefont {M.}~\bibnamefont {Houe}}\ and\ \bibinfo {author} {\bibfnamefont {P.~D.}\ \bibnamefont {Townsend}},\ }\bibfield  {title} {\bibinfo {title} {An introduction to methods of periodic poling for second-harmonic generation},\ }\href {https://doi.org/10.1088/0022-3727/28/9/001} {\bibfield  {journal} {\bibinfo  {journal} {Journal of Physics D: Applied Physics}\ }\textbf {\bibinfo {volume} {28}},\ \bibinfo {pages} {1747} (\bibinfo {year} {1995})}\BibitemShut {NoStop}%
\bibitem [{\citenamefont {Chen}\ \emph {et~al.}(2016)\citenamefont {Chen}, \citenamefont {Karpinski}, \citenamefont {Shvedov}, \citenamefont {Boes}, \citenamefont {Mitchell}, \citenamefont {Krolikowski},\ and\ \citenamefont {Sheng}}]{Chen:16}%
  \BibitemOpen
  \bibfield  {author} {\bibinfo {author} {\bibfnamefont {X.}~\bibnamefont {Chen}}, \bibinfo {author} {\bibfnamefont {P.}~\bibnamefont {Karpinski}}, \bibinfo {author} {\bibfnamefont {V.}~\bibnamefont {Shvedov}}, \bibinfo {author} {\bibfnamefont {A.}~\bibnamefont {Boes}}, \bibinfo {author} {\bibfnamefont {A.}~\bibnamefont {Mitchell}}, \bibinfo {author} {\bibfnamefont {W.}~\bibnamefont {Krolikowski}},\ and\ \bibinfo {author} {\bibfnamefont {Y.}~\bibnamefont {Sheng}},\ }\bibfield  {title} {\bibinfo {title} {Quasi-phase matching via femtosecond laser-induced domain inversion in lithium niobate waveguides},\ }\href {https://doi.org/10.1364/OL.41.002410} {\bibfield  {journal} {\bibinfo  {journal} {Opt. Lett.}\ }\textbf {\bibinfo {volume} {41}},\ \bibinfo {pages} {2410} (\bibinfo {year} {2016})}\BibitemShut {NoStop}%
\bibitem [{\citenamefont {Kibble}\ and\ \citenamefont {Berkshire}(2004)}]{kibble2004classical}%
  \BibitemOpen
  \bibfield  {author} {\bibinfo {author} {\bibfnamefont {T.}~\bibnamefont {Kibble}}\ and\ \bibinfo {author} {\bibfnamefont {F.}~\bibnamefont {Berkshire}},\ }\href {https://books.google.co.il/books?id=0a8dk0eDxgEC} {\emph {\bibinfo {title} {Classical Mechanics}}},\ G - Reference,Information and Interdisciplinary Subjects Series\ (\bibinfo  {publisher} {Imperial College Press},\ \bibinfo {year} {2004})\BibitemShut {NoStop}%
\bibitem [{\citenamefont {Tunyagi}\ \emph {et~al.}(2003)\citenamefont {Tunyagi}, \citenamefont {Ulex},\ and\ \citenamefont {Betzler}}]{PhysRevLett.90.243901}%
  \BibitemOpen
  \bibfield  {author} {\bibinfo {author} {\bibfnamefont {A.~R.}\ \bibnamefont {Tunyagi}}, \bibinfo {author} {\bibfnamefont {M.}~\bibnamefont {Ulex}},\ and\ \bibinfo {author} {\bibfnamefont {K.}~\bibnamefont {Betzler}},\ }\bibfield  {title} {\bibinfo {title} {Noncollinear optical frequency doubling in strontium barium niobate},\ }\href {https://doi.org/10.1103/PhysRevLett.90.243901} {\bibfield  {journal} {\bibinfo  {journal} {Phys. Rev. Lett.}\ }\textbf {\bibinfo {volume} {90}},\ \bibinfo {pages} {243901} (\bibinfo {year} {2003})}\BibitemShut {NoStop}%
\bibitem [{\citenamefont {Baudrier-Raybaut}\ \emph {et~al.}(2004)\citenamefont {Baudrier-Raybaut}, \citenamefont {Ha{\"i}dar}, \citenamefont {Kupecek}, \citenamefont {Lemasson},\ and\ \citenamefont {Rosencher}}]{Baudrier-Raybaut2004}%
  \BibitemOpen
  \bibfield  {author} {\bibinfo {author} {\bibfnamefont {M.}~\bibnamefont {Baudrier-Raybaut}}, \bibinfo {author} {\bibfnamefont {R.}~\bibnamefont {Ha{\"i}dar}}, \bibinfo {author} {\bibfnamefont {P.}~\bibnamefont {Kupecek}}, \bibinfo {author} {\bibfnamefont {P.}~\bibnamefont {Lemasson}},\ and\ \bibinfo {author} {\bibfnamefont {E.}~\bibnamefont {Rosencher}},\ }\bibfield  {title} {\bibinfo {title} {Random quasi-phase-matching in bulk polycrystalline isotropic nonlinear materials},\ }\href {https://doi.org/10.1038/nature03027} {\bibfield  {journal} {\bibinfo  {journal} {Nature}\ }\textbf {\bibinfo {volume} {432}},\ \bibinfo {pages} {374} (\bibinfo {year} {2004})}\BibitemShut {NoStop}%
\bibitem [{\citenamefont {Fischer}\ \emph {et~al.}(2008)\citenamefont {Fischer}, \citenamefont {Saltiel}, \citenamefont {Neshev}, \citenamefont {Krolikowski},\ and\ \citenamefont {Kivshar}}]{Fischer2008}%
  \BibitemOpen
  \bibfield  {author} {\bibinfo {author} {\bibfnamefont {R.}~\bibnamefont {Fischer}}, \bibinfo {author} {\bibfnamefont {S.~M.}\ \bibnamefont {Saltiel}}, \bibinfo {author} {\bibfnamefont {D.~N.}\ \bibnamefont {Neshev}}, \bibinfo {author} {\bibfnamefont {W.}~\bibnamefont {Krolikowski}},\ and\ \bibinfo {author} {\bibfnamefont {Y.~S.}\ \bibnamefont {Kivshar}},\ }\bibfield  {title} {\bibinfo {title} {Transverse second-harmonic generation from disordered nonlinear crystals},\ }\href {https://doi.org/10.2478/s11534-008-0073-6} {\bibfield  {journal} {\bibinfo  {journal} {Central European Journal of Physics}\ }\textbf {\bibinfo {volume} {6}},\ \bibinfo {pages} {569} (\bibinfo {year} {2008})}\BibitemShut {NoStop}%
\bibitem [{\citenamefont {Varon}\ \emph {et~al.}(2011)\citenamefont {Varon}, \citenamefont {Porat},\ and\ \citenamefont {Arie}}]{Varon:11}%
  \BibitemOpen
  \bibfield  {author} {\bibinfo {author} {\bibfnamefont {I.}~\bibnamefont {Varon}}, \bibinfo {author} {\bibfnamefont {G.}~\bibnamefont {Porat}},\ and\ \bibinfo {author} {\bibfnamefont {A.}~\bibnamefont {Arie}},\ }\bibfield  {title} {\bibinfo {title} {Controlling the disorder properties of quadratic nonlinear photonic crystals},\ }\href {https://doi.org/10.1364/OL.36.003978} {\bibfield  {journal} {\bibinfo  {journal} {Opt. Lett.}\ }\textbf {\bibinfo {volume} {36}},\ \bibinfo {pages} {3978} (\bibinfo {year} {2011})}\BibitemShut {NoStop}%
\end{thebibliography}%
\clearpage
\appendix
\section{End Matter}
\textit{Appendix A: 1D transverse localization of the idler-signal pseudospin}—Creating a 2D disordered transverse synthetic magnetization pattern through crystal shaping requires 3D-modulated NLPCs. However, current fabrication techniques are limited to crystal lengths shorter than the propagation distances needed to observe 2D transverse localization in our system \cite{Zhang:21, Imbrock:20, Yu:24}, restricting experimental exploration. One alternative approach is to use naturally disordered crystals \cite{PhysRevLett.90.243901, Baudrier-Raybaut2004, Fischer2008}, though they lack control over disorder properties. A more feasible experimental approach is to introduce disorder along a single transverse axis ($\hat{\textbf{x}}$ or $\hat{\textbf{y}}$), requiring only 2D-modulated NLPCs \cite{Varon:11}. This method effectively creates a 1D optical spin-glass phase, enabling the observation of 1D transverse localization of the idler-signal pseudospin.
\begin{figure}[ht]
\includegraphics[width=\linewidth]{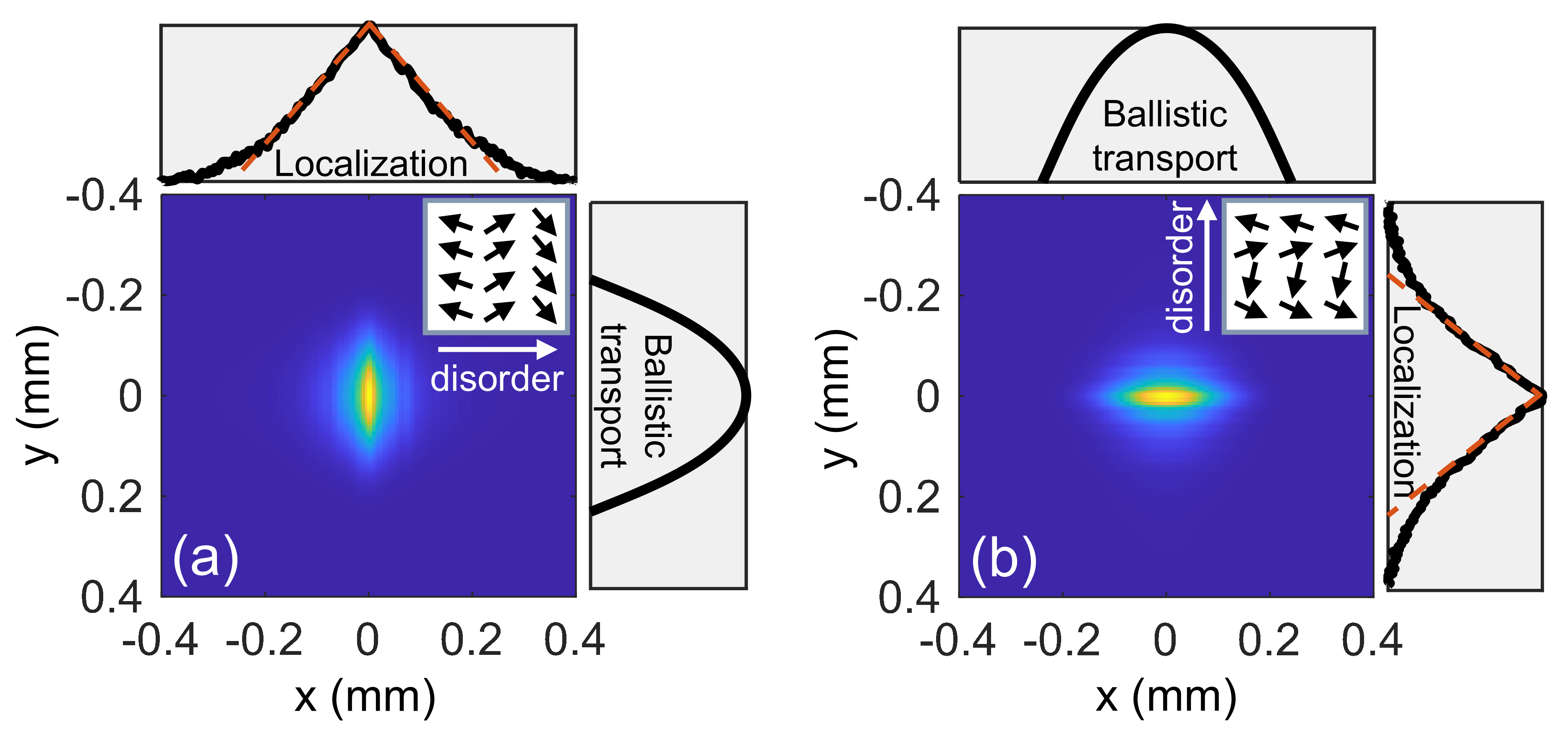}
\caption{\label{fig:figure5}\textbf{1D transverse localization in a synthetically disordered 2D-modulated NLPC for a pump intensity of 142.5 MW/cm$^{2}$.} (a) and (b) show the ensemble-averaged idler-signal output intensity cross-sections for disorder along the $\hat{\textbf{x}}$ or $\hat{\textbf{y}}$ axes, respectively, with insets illustrating the corresponding synthetic magnetization patterns. The black curves above panels (a) and (b) depict logarithmic intensity profiles along the $\hat{\textbf{x}}$-axis, while those to the right show profiles along the $\hat{\textbf{y}}$-axis. The localization length of the ensemble-averaged idler-signal output along the disordered $\hat{\textbf{x}}$-axis ($\hat{\textbf{y}}$-axis) is approximately $\xi_{\text{loc}}^x = 80 \ \mu m$ ($\xi_{\text{loc}}^y = 75 \ \mu m$), with a corresponding full width at half maximum (FWHM) of about $55 \ \mu m$ ($52 \ \mu m$), three times smaller than the FWHM of $167 \ \mu m$ ($166 \ \mu m$) along the ordered axis.}
\end{figure}

In this approach, the modulation phase, $\phi_{\text{NLPC}}(\textbf{r})$, is uniformly sampled from $\left[0,2\pi\right]$ along one transverse axis, while remaining constant along the other transverse axis (as well as along the propagation axis). This is illustrated in the insets of Figs.~\ref{fig:figure5}(a) and ~\ref{fig:figure5}(b). As a result, the ensemble-averaged idler-signal output intensity becomes localized along the disordered axis, exhibiting exponential decaying tails, while broadening diffractively along the ordered axis. Figs.~\ref{fig:figure5}(a) and ~\ref{fig:figure5}(b) show this behavior for disorder introduced along the $\hat{\textbf{x}}$ and $\hat{\textbf{y}}$ axes, respectively. The logarithmic intensity profiles further emphasize this phenomenon, showing a linear curve along the disordered axis, indicating 1D localization, and a parabolic curve along the ordered axis, signifying ballistic transport.

\textit{Appendix B: Idler-signal pseudospin decoherence effect}—The analogy between an SFG process and spin-1/2 dynamics can be geometrically visualized on a frequency-domain Bloch sphere \cite{PhysRevA.78.063821, https://doi.org/10.1002/lpor.201300107, Karnieli2021_FoP}. In this representation, the effective pseudospin state vector, $\Psi(\textbf{r}) \propto \left ( A_{i}(\textbf{r}), A_{s}(\textbf{r}) \right )^{T}$, is mapped onto the Bloch sphere, with the idler and signal fields corresponding to the north and south poles, respectively \cite{PhysRevA.78.063821, Karnieli2021_FoP}. Equal superpositions of the idler and signal fields lie along the equator, while other points on the sphere represent states with varying idler-signal populations and relative phases. The nonlinear interaction between the idler and signal fields can be described by the precession of the Bloch vector, $\textbf{S} = \Psi^\dagger \pmb{\sigma} \Psi$, around the synthetic magnetization vector $\textbf{M}(\textbf{r})$ \cite{PhysRevA.78.063821, Karnieli2021_FoP}:
\begin{equation}
    \frac{\partial}{\partial z} \textbf{S} = \textbf{M} \times \textbf{S}.
\end{equation}

Here, $\textbf{M}(\textbf{r}) = M_0(\textbf{r}) \hat{\textbf{M}}_T(\phi(\textbf{r})) + \frac{\Delta k}{2} \hat{\textbf{z}}$, where $\Delta k = k_p + k_i - k_s - k_c$ is the overall phase mismatch of the process, owing to the wavenumber difference of the interacting fields and the crystal's effective momentum, $k_{c}$. The Bloch components in terms of field amplitudes are: $S_x = \frac{1}{N} \sqrt{\frac{n_i n_s}{\omega_i \omega_s}} 2 \operatorname{Re}\left( e^{-i \Delta k z /2} A_i^* A_s \right)$, $S_y = \frac{1}{N} \sqrt{\frac{n_i n_s}{\omega_i \omega_s}} 2 \operatorname{Im}\left( e^{-i \Delta k z / 2} A_i^* A_s \right)$ and $S_z = \frac{1}{N} \left( \frac{n_i}{\omega_i} |A_i|^2 - \frac{n_s}{\omega_s} |A_s|^2 \right)$. The magnitude of the Bloch vector, $|\textbf{S}| = \sqrt{S_x^2 + S_y^2 + S_z^2}$, is 1 for a pure state and less than 1 for a mixed state. In our analysis, we assume a quasi-phase-matched process where $\textbf{M}(\textbf{r}) = \textbf{M}_{T}(\textbf{r})$ lies on the equator of the Bloch sphere, and an initial pseudospin state vector of $\Psi(\textbf{r}) \propto A_i(\textbf{r}) \otimes | \omega_i \rangle$, causing the initial Bloch vector to point to the north pole. 

\begin{figure}[ht]
\includegraphics[width=\linewidth]{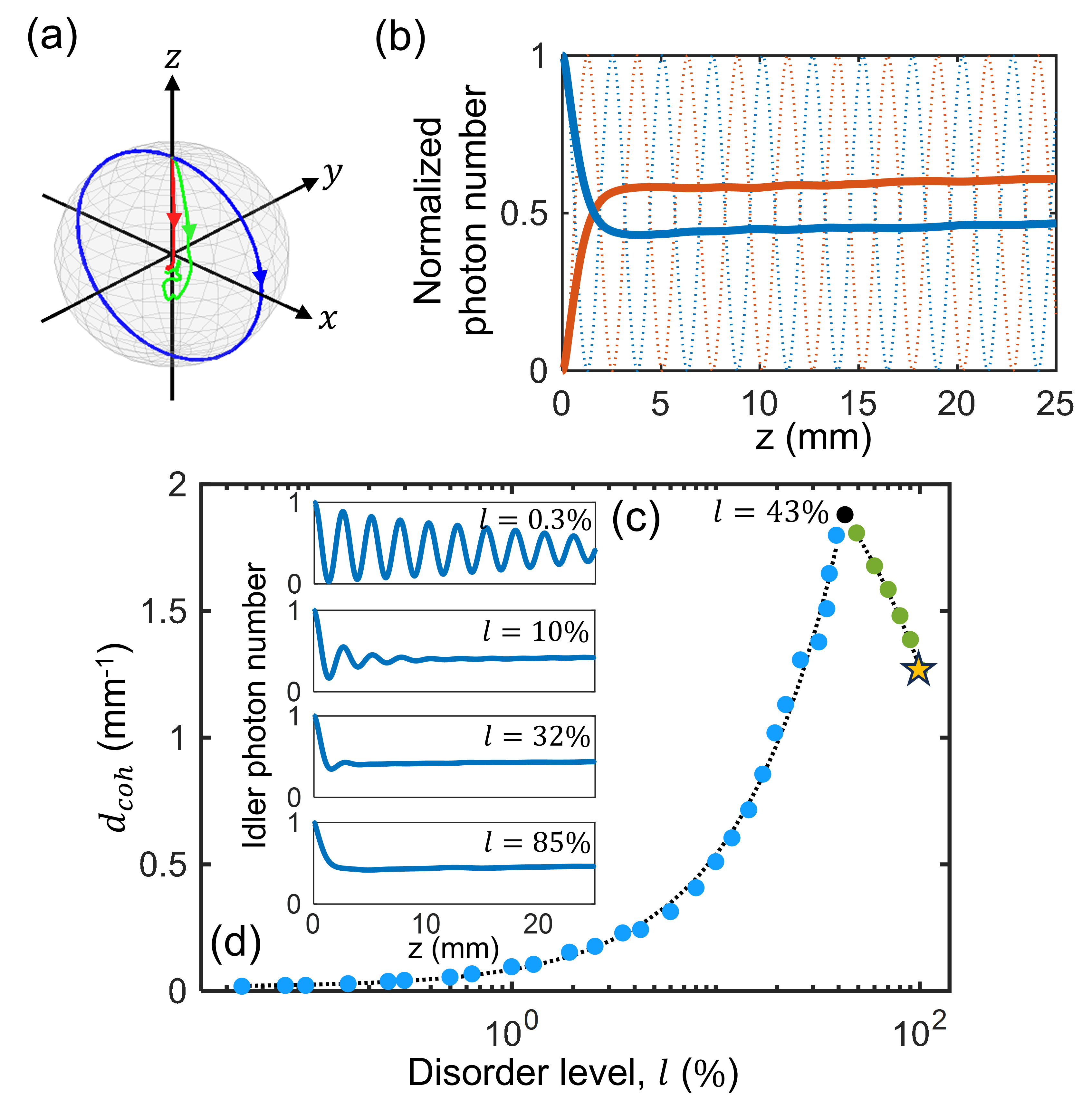}
\caption{\label{fig:figure6}\textbf{Pseudospin-decoherence effect in a synthetically disordered 3D-modulated NLPC for a pump intensity of 142.5 MW/cm$^{2}$.} (a) Trajectories of the system's Bloch vector, $\textbf{S}$, on the frequency-domain Bloch sphere. The blue trajectory represents the ordered ferromagnetic system $\left(l=0\%\right)$, while the green and red trajectories correspond to the disordered spin-glass system with $l=100\%$, for a single disorder realization and an ensemble average, respectively. (b) Normalized photon numbers of the idler (blue) and signal (red) fields as functions of propagation distance, $z$. Dashed curves depict Rabi oscillations in the ordered system $\left(l=0\%\right)$, while solid curves show overdamped oscillations in the disordered system with $l=100\%$. (c) Normalized idler photon number versus propagation distance for increasing disorder levels. (d) The decay coefficient $d_{\text{coh}}$, as a function of disorder level. The black data point at $\left(l=43\%\right)$ marks the transition from underdamped (blue data points) to overdamped (green data points) oscillations, where critically damped behavior is observed. Dashed black curves represent fits for the underdamped and overdamped regimes, and the star symbol denotes the disorder level $\left(l=100\%\right)$ corresponding to the dynamics shown in panel (b).}
\end{figure}

In an optical ferromagnetic system, where $\textbf{M}_{T}$ is uniform, the Bloch vector, $\textbf{S}$, precesses in a circular path around $\textbf{M}_{T}$, oscillating between the north and south poles \cite{PhysRevA.78.063821, Karnieli2021_FoP}, as shown by the blue trajectory in Fig. 6(a). These oscillations, depicted by the dashed curves in Fig.~\ref{fig:figure6}(b), are the "Rabi" oscillations, resulting in complete frequency conversion between the idler and signal fields every half-cycle, depending on the nonlinear interaction parameters \cite{PhysRevA.78.063821, Karnieli2021_FoP}.

In contrast, in the optical disordered spin-glass, where pseudospin transverse localization occurs, our simulations reveal a pseudospin decoherence effect, characterized by decaying Rabi oscillations rather than a full frequency conversion. The initial idler field evolves into an approximately equal mixture of idler and signal fields, as shown by the solid curves in Fig. ~\ref{fig:figure6}(b). Here, the Bloch vector lies inside the Bloch sphere, for both a single disorder realization and an ensemble average, as depicted by the green and red trajectories in Fig. ~\ref{fig:figure6}(a). This behavior indicates that the system transitions into a mixed state after the interaction.

Interestingly, the oscillation decay profile varies with the disorder strength: as the disorder increases, the oscillations shift from underdamped to overdamped, as shown in Fig. ~\ref{fig:figure6}(c). To further investigate this behavior, we calculated the decay coefficient, $d_{\text{coh}}$, for various disorder levels at a fixed pump intensity. As seen in Fig. ~\ref{fig:figure6}(d), a transition from underdamped to overdamped oscillations occurs at a disorder level of $l \approx 43\%$, revealing a clear relationship between the disorder strength and the decay coefficient. This relationship can be expressed in a closed-form equation, enabling us to predict the energy exchange dynamics between the idler and signal fields for a given disorder level.

This longitudinal decoherence results from the scattering of the idler-signal pseudospin by random magnetization domains as it propagates through the synthetically disordered NLPC. After a certain propagation distance, the pseudospin encounters new, statistically independent random domains, leading to coherence loss. This suggests the presence of two interrelated types of disorder in the system: transverse disorder, which drives 2D pseudospin localization, and an effective longitudinal disorder which causes longitudinal decoherence, analogous to temporal decoherence in the spintronic system. As shown in Fig.~\ref{fig:figure6}(d), the synthetic transverse disorder establishes a longitudinal disorder, which in turn governs the decoherence dynamics.

Further discussion on this decoherence effect and the calculation of the oscillations decay coefficient, can be found in SM \cite{supplemental}.

\newpage
\end{document}